\documentclass[12pt,draftcls,peerreview,onecolumn]{IEEEtran}

\usepackage{amsmath,amssymb,graphicx,epstopdf,cite,subfigure}
\usepackage{psfrag}
\usepackage{amssymb}
\usepackage{amsmath}
\usepackage{bbm}
\usepackage{dsfont}
\usepackage{pifont}
\usepackage{cite}
\usepackage{graphics}
\usepackage{graphicx}
\usepackage{epsfig}
\usepackage{subfigure}
\usepackage{amscd}
\usepackage{accents}

\newlength{\dhatheight}

\begin{document}
%
\title{Statistical Recovery of Simultaneously Sparse Time-Varying Signals from Multiple Measurement Vectors }
%
%
%

\author{Jun Won Choi,~\IEEEmembership{Member,~IEEE,} and Byonghyo Shim,~\IEEEmembership{Senior Member,~IEEE}
\thanks{Jun Won Choi is with Dept. of Electrical Engineering, Hanyang University, Seoul Korea, e-mail: junwchoi@hanyang.ac.kr.}
\thanks{Byonghyo Shim is with Dept. of Electrical Engineering, Seoul National University, Seoul Korea, e-mail:
bshim@snu.ac.kr.}
}

\maketitle


\begin{abstract}
In this paper, we propose a new sparse signal recovery algorithm, referred to as sparse Kalman tree search (sKTS), that provides a robust reconstruction of the sparse vector when the sequence of correlated observation vectors are available.
The proposed sKTS algorithm builds on expectation-maximization (EM) algorithm and consists of two main operations: 1) Kalman smoothing to obtain the \emph{a posteriori} statistics of the source signal vectors and 2) greedy tree search to estimate the support of the signal vectors.
Through numerical experiments, we demonstrate that the proposed sKTS algorithm is effective in recovering the sparse signals and  performs close to the Oracle (genie-based) Kalman estimator.
\end{abstract}

\begin{IEEEkeywords}
Compressed sensing, simultaneously sparse signal, multiple measurement vector, expectation-maximization (EM) algorithm, maximum likelihood estimation
\end{IEEEkeywords}

%
\IEEEpeerreviewmaketitle

\section{Introduction}
\label{sec:intro}

Over the years, there has been growing interest in the recovery of high dimensional signals from a small number of measurements. This new paradigm, so called compressed sensing (CS), relies on the fact that many naturally acquired high dimensional signals inherently have low dimensional structure.
%
%
In fact, since many real world signals can be well approximated as sparse signals (i.e., only a few entries of signal vector are nonzero), CS techniques have been applied to a variety of applications including data compression, source localization, wireless sensor network, medical imaging, data mining, to name just a few.
%

Over the years, various signal recovery algorithms for CS have been proposed. Roughly speaking, these approaches are categorized into two classes. The first approach is based on a deterministic signal model, where an underlying signal is seen as a deterministic vector and the sparsity promoting cost function (e.g., $\ell_1$-norm) is employed to solve the problem.  These approaches include the basis pursuit (BP) \cite{candes}, orthogonal matching pursuit (OMP) \cite{omp,gomp}, CoSaMP \cite{cosamp}, and subspace pursuit \cite{wei}.
The second approach is based on the probabilistic signal model, where the signal sparsity is described by the \emph{a priori} distribution of the signal and Bayesian framework is employed in finding the sparse solution \cite{rao_basis,bcs}.

When the multiple measurement vectors (MMV) from different source signals with common support are available, accuracy of the sparse signal recovery can be improved dramatically by performing joint processing of these vectors \cite{tropp,rao_mmv,tropp2,rao_emp,rao_temporal,rao_ar,prasad}. Since the algorithms based on MMV usually performs better than those relying on single measurement vector, many efforts have been made in recent years to develop an efficient sparse recovery algorithm.
The MMV-based recovery algorithms targeted for the deterministic signal recovery include the mixed-norm solution \cite{tropp,rao_mmv} and convex relaxation \cite{tropp2} while the probabilistic approaches include the MMV sparse Bayesian learning (SBL) method \cite{rao_emp}, block-SBL \cite{rao_temporal}, auto-regressive SBL \cite{rao_ar}, and Kalman filtering-based SBL (KSBL) \cite{prasad}.

In this work, we are primarily concerned with the MMV-based signal recovery problem when the observation vectors are sequentially acquired. To be specific, we express the $N \times 1$ observation vector $\mathbf{y}_{n}$ acquired at time index $n$ as
\begin{align} \label{eq:smodel}
 \mathbf{y}_{n}= \mathbf{B}_{n} \mathbf{h}_{n} + \mathbf{w}_{n},
\end{align}
where $\mathbf{B}_{n}$ is the $N \times M$ system matrix,  $\mathbf{h}_{n}$ is the $M \times 1$ source signal vector, and $\mathbf{w}_{n}$ are the $N \times 1$ noise vector. We assume that $\mathbf{w}_{n}$ is modeled as a zero mean complex Gaussian random vector, i.e., $\mathcal{CN}(0,\sigma_{w}^{2}\mathbf{I})$.
%
%
Our goal in this setup is to estimate the source signal $\mathbf{h}_{n}$ using the sequence of the observations $\{\mathbf{y}_{n} \}$ when 1) the source signal $\mathbf{h}_{n}$ is sparse (i.e., the number nonzero elements in $\mathbf{h}_{n}$ is small) and 2) the dimension of the observation vector $\mathbf{y}_{n}$ is smaller than that of the source vector  $(N \ll M)$.
In particular, we  focus on the scenario where  the nonzero elements of $\mathbf{h}_{n}$ change over time with certain temporal correlations. In this scenario, we assume that correlated time-varying signals are well modeled by Gauss-Markov process. Note that this model is useful in capturing local dynamics of signals in linear estimation theory \cite{kailath}.

\begin{figure} [t]
 \centering
\centerline{\epsfig{figure=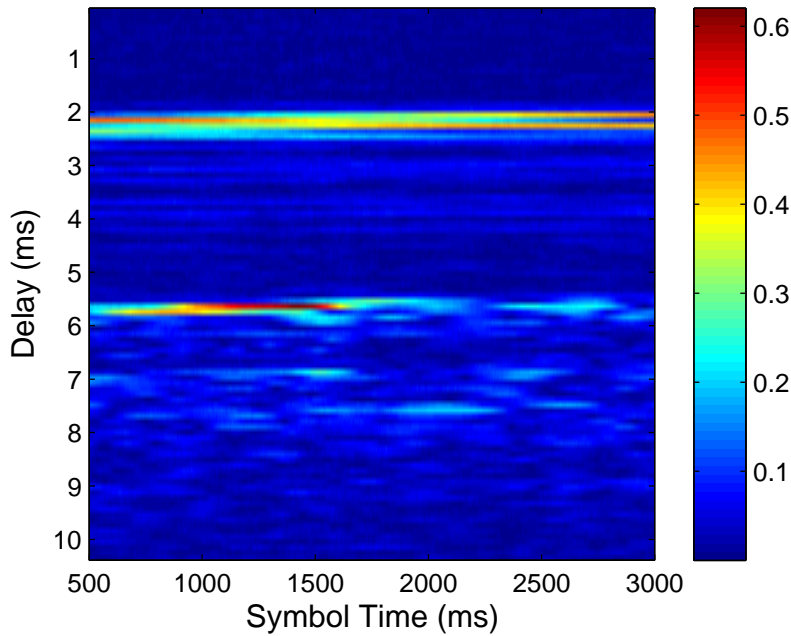, width=120mm}}
  \caption {A record of the channel impulse response of underwater acoustic channels measured off the coast of Martha's Vinyard, MA, USA.}
  \label{fig:cir}
\end{figure}

The main purpose of this paper is to propose a new statistical sparse signal estimation algorithm for the sequential observation model we just described. The underlying assumption used in our model is that the nonzero amplitude of the sparse signals is changing in time, leading to different signal realizations for each measurement vector, yet the support of the signal amplitude  is slowly varying so that the support remains unchanged over certain consecutive measurement vectors.  We henceforth refer to this model as {\it simultaneously sparse signal with locally common support} since the support of the sparse signal remains constant over the fixed interval under this assumption.
Many of signal processing and wireless communication systems are characterized by this model.
%
For example, this model matches well with the characteristics of multi-path fading channels for wireless communications where the channel impulse response $\mathbf{h}_{n}$ should be estimated from the received signal $\mathbf{y}_{n}$. Fig. \ref{fig:cir} shows a record of the channel impulse responses (CIR) of underwater acoustic channels (represented over the propagation delay and time domain) measured from the experiments conducted in Atlantic ocean in USA \cite{jchoi}. We observe that when compared to the amplitude of the channel taps, the sparsity structure of the CIR is varying slowly. Thus, we can readily characterize this time-varying sparse signal using the correlated random process along with a deterministic binary parameter representing the existence of the signal.

In recovering the original signal vector $\mathbf{h}_{n}$ from the measurement vectors, we use the modified expectation-maximization (EM) algorithm \cite{em}. The proposed scheme, dubbed as sparse-Kalman-Tree-Search (sKTS), consists of two main operations: 1) Kalman smoothing to gather the \emph{a posteriori} statistics of the source signals from individual measurement vector within the block of interest and 2) identification of the support of the sparse signal vector using a greedy tree search algorithm. Treating the problem to identify the sparsity structure of the source signal as a combinatorial search, we propose a simple yet effective greedy tree search algorithm that examines the small number of promising candidates among all sparsity parameter vectors in the tree.

There exist several approaches to estimate the time-varying sparse signals under MMV model. In  \cite{charles1, charles2}, reweighted $\ell_1$ optimization has been modified for the sequential dynamic filtering. In \cite{prasad}, modified SBL algorithm has been suggested to adopt autoregressive modeling.
In \cite{nicholas}, EM-based adaptive filtering scheme has been proposed in the context of sparse channel estimation. Other than these, notable approaches include turbo approximate message passing (AMP) \cite{schniter2}, Lasso-Kalman \cite{giannakis}, and Kalman filtered CS \cite{kalmancs}. We note that our work is distinct from these approaches in the following two aspects.
First, in contrast to the previous efforts using continuous (real-valued) parameters to describe signal sparsity  in \cite{charles1, charles2,prasad},  the proposed method employs the deterministic discrete (binary) parameter vector that captures the on-off structure of signal sparsity. Due to the use of deterministic parameter vector, an effort to deal with the probabilistic model on signal sparsity is unnecessary. Also, since the search space is discretized, identification of parameter vector is done by the efficient search algorithm.
Second, while the recent work in \cite{kalmancs} estimates signal amplitude using Kalman smoother and then identifies the support of sparse signal by thresholding of the innovation error norm, our work pursues direct estimation of the binary parameter vector using the modified EM algorithm.
%
We note that a part of this paper was presented in \cite{doubly}. The distinctive contribution of the present work is that the algorithm is developed in a more generic system model and practical issues (e.g., parameter estimation and iteration control) and real-time implementation issues are elaborated. Further,  extensive simulations for the practical applications are conducted to demonstrate the superiority of the proposed method.


The rest of this paper is organized as follows.
In Section \ref{sec:sKTS_proposed}, we briefly explain the sparse signal model and then present the proposed method.
In Section \ref{sec:chan}, we discuss the application of the proposed algorithm in the wireless channel estimation.
In section \ref{sec:simul}, the simulation results are provided, and Section \ref{sec:conclusion} concludes the paper.

Notation:
Uppercase and lowercase letters written in boldface denote matrices and vectors, respectively.
Superscripts $(\cdot)^{T}$ and $(\cdot)^{H}$ denote  transpose and  conjugate transpose (hermitian operator), respectively.
${\rm conj}(x)$ denotes the conjugation of the complex number $x$.
$\| \cdot\|_{p}$ indicates an $\ell_p$-norm of a vector.
For the $\ell_2$-norm, we abbreviate a subscript $p$ for simplicity.
%
%
${\rm diag}\left(\cdot\right)$ is a diagonal matrix having elements only on the main diagonal.
${\rm Re}\left(x\right)$ and ${\rm Im}\left(x\right)$ denote the real and imaginary parts of $x$, respectively.
$E[X]$ denotes the expectation of a random variable $X$ and $E[X|Y]$ denotes the conditional expectation of $X$ given $Y$.
$E[X;\theta]$ means the expectation of $X$ given the deterministic parameter $\theta$.
The notations for covariance matrices are given by ${\rm Cov}(\mathbf{x},\mathbf{y}) = E[\mathbf{x}\mathbf{y}^{H}] - E[\mathbf{x}]E[\mathbf{y}]^{H}$ and
${\rm Cov}(\mathbf{x}) = {\rm Cov}(\mathbf{x},\mathbf{x}) $.
%
%
%
$Pr(A)$ means the probability of the event $A$.  ${\rm tr} (A)$ denotes a trace operation of the matrix $A$.
$A \odot B$ is the element-by-element product (Hadamard product) of the matrices $A$ and $B$.
$\mathbf{e}_{i}$ denotes the $i$th coordinate vector.

\section{Proposed Sparse Signal Estimation Technique}
\label{sec:sKTS_proposed}

In this section, we consider the statistical estimation of the time-varying sparse signals
from the sequentially collected observation vectors. As mentioned, our approach is based on the assumption that the support of the sparse signal varies slowly in time so that the multiple measurement vectors sharing common support can be used to improve the estimation quality of the sparse signals.  In this section, we first describe the simultaneously sparse signal model   and then present the proposed sparse signal estimation scheme.

\subsection{Simultaneously Sparse Signal Model}
\label{sec:sKTS_block}

We express a time-varying sparse signal $\mathbf{h}_{n}$ as a product of a vector of random processes  $\mathbf{s}_{n}$ describing the amplitudes  of nonzero entries in $\mathbf{h}_{n}$ and the vector  $\mathbf{c}_{i} = [c_{i,1}, \cdots, c_{i,M}]^{T}$ indicating the existence of signal. That is,
\begin{align}
\mathbf{h}_n &= {\rm diag}(\mathbf{c}_{i}) \mathbf{s}_{n},
\end{align}
where $i$ is the block index, the entry of $\mathbf{c}_{i}$ is either 0 or 1 depending on the existence of the signal
\begin{align}
c_{i,j} = \left\{ \begin{array}{cc}
                         1 & \mbox{if the $j$th entry of $\mathbf{h}_n$ exists} \\
                         0 & \mbox{otherwise,}
                       \end{array}
\right.
\end{align}
and the time-varying amplitude $\mathbf{s}_{n}  \in \mathcal{C}^{M}$ is modeled as Gauss-Markov random process
\begin{align}
\mathbf{s}_{n+1} &= \mathbf{A}_{n}\mathbf{s}_{n} + \mathbf{v}_{n},
\end{align}
where  $\mathbf{v}_{n} \in \mathcal{C}^{M}$ is the process noise vector ($\sim \mathcal{CN}(0,\mathbf{V}_{n})$)  and $\mathbf{A}_{n} \in \mathcal{C}^{M \times M}$ is the state update matrix. Note that the block index $i$ is associated with the interval of the length $T$, $n \in [Ti+1, T(i+1)]$.
As mentioned, we assume that the support of the underlying sparse signals is locally time-invariant so that  $\mathbf{c}_{i}$  is constant in a block of consecutive measurement vectors.
Using this together with the observation model in (\ref{eq:smodel}), we obtain the simultaneously sparse signal model
\begin{align}
\mathbf{s}_{n+1} &= \mathbf{A}_{n}\mathbf{s}_{n} + \mathbf{v}_{n},  \nonumber \\
\mathbf{h}_n &= {\rm diag}(\mathbf{c}_{i}) \mathbf{s}_{n} \nonumber \\
\mathbf{y}_{n} &= \mathbf{B}_{n} \mathbf{h}_{n} +  \mathbf{w}_{n}.
 \label{eq:blocks}
\end{align}
Since $\mathbf{h}_{n}$ follows Gaussian distribution for the given $\mathbf{c}_{i}$,  the {\it a priori} distribution of the source signal $\mathbf{h}_{n}$ can be described by
\begin{align}
Pr\left(\mathbf{h}_{n}; \mathbf{c}_{i} \right) = \frac{1}{(2\pi)^{M}{\rm det}({\rm Cov}\left(\mathbf{h}_n \right))}{\rm exp}\left( -(\mathbf{h}_{n}-E\left[\mathbf{h}_n; \mathbf{c}_{i} \right])^{H}{\rm Cov}\left(\mathbf{h}_n \right)^{-1} (\mathbf{h}_{n}-E\left[\mathbf{h}_n; \mathbf{c}_{i} \right]) \right),
\end{align}
where
\begin{align}
E\left[\mathbf{h}_n;  \mathbf{c}_{i}  \right] &= {\rm diag}(\mathbf{c}_{i}) E[\mathbf{s}_{n}] \nonumber \\
{\rm Cov}\left(\mathbf{h}_n \right) &= {\rm diag}(\mathbf{c}_{i}) {\rm Cov} (\mathbf{s}_{n}){\rm diag}(\mathbf{c}_{i}). \label{eq:sttt}
\end{align}

\subsection{Derivation of Statistical Sparse Signal Estimation}
\label{sec:sKTS_dv}

When the multiple measurement vectors $\{\mathbf{y}_{Ti+1}, \cdots, \mathbf{y}_{T(i+1)}\}$ in the $i$th block are available,  the maximum likelihood (ML)  estimate of $\mathbf{c}_{i}$ is expressed as
\begin{align} \label{eq:mlsol}
\mathbf{c}_{i}^{\rm ML} = \arg \max_{\mathbf{c}_{i} \in \{0,1\}^T, \sum_{j=1}^{M}c_{i,j}=K} \ln \Pr\left(\mathbf{y}_{1:T} ; \mathbf{c}_{i}\right),
\end{align}
where $\mathbf{y}_{1:T}=[\mathbf{y}_{Ti+1}, \cdots, \mathbf{y}_{T(i+1)}]^{T}$ and  $K$ is the sparsity order (the number of nonzero entries) of $\mathbf{h}_{n}$. Note that the subscript $1:T$ denotes the set of time indices for the $i$th block. Note also that the ML estimate $\mathbf{c}_{i}^{\rm ML}$ is chosen among all candidates satisfying the sparsity constraint  $\left( \sum_{j=1}^{M}c_{i,j}=K \right)$.
Once $\mathbf{c}_{i}^{\rm ML}$ is obtained, we can estimate the amplitude vectors $\mathbf{s}_{n}$ assuming that the signal support specified by $\mathbf{c}_{i}^{\rm ML}$ is true. Well known linear minimum mean square error (LMMSE) estimator  (e.g. Kalman smoother) can be used to estimate $\mathbf{s}_{n}$ and then  $\mathbf{c}_{i}^{\rm ML}$ and the estimate of $\mathbf{s}_{n}$ are combined to produce a final estimate of $\mathbf{h}_{n}$. Note that if the estimation of $\mathbf{c}_{i}$ is correct, we can obtain the best achievable estimate of $\mathbf{h}_{n}$, which is equivalent to the solution attainable by so called  ``Oracle estimator".
Since the ML problem in (\ref{eq:mlsol}) involves the marginalization over all possible combinations of the latent variables $\mathbf{s}_{1:T}$, finding out the solution using the direct approach would be computationally unmanageable.
Perhaps, a better way to deal with the problem at hand is to use the EM algorithm. Recall that the EM algorithm is an efficient means to find out the ML estimate or maximum a posteriori (MAP) estimate of statistical signal model in the presence of unobserved latent variables.
 The EM algorithm generates a sequence of estimates $\hat{\mathbf{c}}_i^{(l)}$, $l=0,1,2,...$ by alternating two major steps (E-step and M-step), which are given, respectively
\begin{itemize}
\item \textbf{Expectation step (E-step)}
\begin{align} \label{eq:estep}
Q\left(\mathbf{c}_{i}; \hat{\mathbf{c}}_i^{(l)}\right) = E\left[ \ln \Pr\left(\mathbf{y}_{1:T},\mathbf{s}_{1:T} ; \mathbf{c}_{i}\right) \bigg|  \mathbf{y}_{1:T} ; \hat{\mathbf{c}}_i^{(l)}\right],
\end{align}
\item \textbf{Maximization step (M-step)}
\begin{align} \label{eq:devm_step}
\hat{\mathbf{c}}_i^{(l+1)} = \arg \max_{\mathbf{c}_{i} \in \{0,1\}^M, \sum_{j=1}^{M}c_{i,j}=K} Q\left(\mathbf{c}_{i}; \hat{\mathbf{c}}_i^{(l)}\right) ,
\end{align}
\end{itemize}
where $\hat{\mathbf{c}}_i^{(l)}$ is the estimate of $\mathbf{c}_{i}$ at the $l$-th iteration.
%
%
Although one cannot guarantee finding out the global optimal solution of (\ref{eq:mlsol}) using the EM algorithm,
we will empirically show that $\mathbf{c}_{i}$ can be estimated accurately with a proper initialization of $\mathbf{c}_{i}^{(0)}$ (see Section \ref{sec:simul}).

\subsection{\textbf{The E-step}}
\label{sec:estep}
The goal of the E-step is to obtain a simple expression of the cost metric $Q(\mathbf{c}_{i},\hat{\mathbf{c}}_i^{(l)})$ using the simultaneously sparse signal model. First, $\ln \Pr\left(\mathbf{y}_{1:T},\mathbf{s}_{1:T} ; \mathbf{c}_{i}\right)$ is expressed as
\begin{align}
\ln \Pr\left(\mathbf{y}_{1:T},\mathbf{s}_{1:T} ; \mathbf{c}_{i}\right) &= \ln \Pr\left(\mathbf{y}_{1:T}|\mathbf{s}_{1:T} ; \mathbf{c}_{i}\right) +  \ln \Pr\left(\mathbf{s}_{1:T} ; \mathbf{c}_{i}\right) \\
=& \sum_{n=Ti+1}^{T(i+1)} \ln \Pr\left(\mathbf{y}_{n}|\mathbf{s}_{n} ; \mathbf{c}_{i}\right) +  \sum_{n=Ti+1}^{T(i+1)} \ln \Pr\left(\mathbf{s}_{n}|\mathbf{s}_{n-1} \right)
\end{align}
Noting that
$\Pr\left(\mathbf{y}_{n}|\mathbf{s}_{n} ; \mathbf{c}_{i}\right) \sim {\mathcal CN}(\mathbf{B}_{n}{\rm diag}(\mathbf{c}_{i}) \mathbf{s}_{n}, \sigma_{w}^{2}\mathbf{I})$ and $\Pr\left(\mathbf{s}_{n}|\mathbf{s}_{n-1} \right) \sim {\mathcal CN}(
\mathbf{A}_{n-1}\mathbf{s}_{n-1}, \mathbf{V}_{n-1})$, we have
\begin{align}
\ln \Pr\left(\mathbf{y}_{1:T},\mathbf{s}_{1:T} ; \mathbf{c}_{i}\right) =& - \sum_{n=Ti+1}^{T(i+1)} \frac{1}{\sigma_{w}^{2}} \left\|\mathbf{y}_{n} - \mathbf{B}_{n}{\rm diag}(\mathbf{c}_{i}) \mathbf{s}_{n} \right\|^{2} \nonumber \\  \label{eq:vv}
& \;\;\;\;\;\; -  \sum_{n=Ti+1}^{T(i+1)} \left(\mathbf{s}_{n} - \mathbf{A}_{n-1}\mathbf{s}_{n-1}\right)^{H}\mathbf{V}_{n-1}^{-1} \left(\mathbf{s}_{n} - \mathbf{A}_{n-1}\mathbf{s}_{n-1}\right) + C
\\
=&  - \sum_{n=Ti+1}^{T(i+1)} \frac{1}{\sigma_{w}^{2}} \left\|\mathbf{y}_{n} - \mathbf{B}_{n}{\rm diag}(\mathbf{c}_{i}) \mathbf{s}_{n} \right\|^{2} + C', \label{eq:vv2}
\end{align}
where $C$ and $C'$ are the terms independent of  $\mathbf{c}_{i}$.
From (\ref{eq:estep}) and (\ref{eq:vv2}), we further have (see Appendix \ref{appen:q})
\begin{align}
Q\left(\mathbf{c}_{i}; \hat{\mathbf{c}}_i^{(l)}\right) =&
 C'' + \frac{1}{\sigma_{w}^{2}} \sum_{n=Ti+1}^{T(i+1)} \bigg\{    2 {\rm Re}\left(  \mathbf{y}_{n}^{H} \mathbf{B}_{n} {\rm diag}(\mathbf{c}_{i}) E\left[\mathbf{s}_{n}\bigg| \mathbf{y}_{1:T} ; \hat{\mathbf{c}}_i^{(l)} \right] \right)  \nonumber \\
& -{\rm tr}\left[  \mathbf{B}_{n} {\rm diag}(\mathbf{c}_{i})  E\left[ \mathbf{s}_{n}\mathbf{s}_{n}^{H}\bigg|  \mathbf{y}_{1:T} ; \hat{\mathbf{c}}_i^{(l)} \right] {\rm diag}(\mathbf{c}_{i})  \mathbf{B}_{n}^{H}  \right]   \bigg\}. \label{eq:qqq}
\end{align}
 Let $\widehat{\mathbf{s}}_{n|1:T}$ and  $\Sigma_{n|1:T}$ be the conditional mean and covariance of $\mathbf{s}_{n}$ when $\mathbf{y}_{1:T} $ and $\hat{\mathbf{c}}_i^{(l)}$ are given, i.e.,
  \begin{align}
  \widehat{\mathbf{s}}_{n|1:T} &=  E\left[\mathbf{s}_{n}\bigg|  \mathbf{y}_{1:T} ; \hat{\mathbf{c}}_i^{(l)} \right] \nonumber \\
  \Sigma_{n|1:T} &= {\rm Cov}\left[ \mathbf{s}_{n}\bigg|  \mathbf{y}_{1:T} ; \hat{\mathbf{c}}_i^{(l)} \right]. \nonumber
  \end{align}
Now We turn to the estimation of the {\it a posteriori} statistics $\widehat{\mathbf{s}}_{n|1:T}$ and $ \Sigma_{n|1:T}$.
In our work, we estimate $\widehat{\mathbf{s}}_{n|1:T}$ and $ \Sigma_{n|1:T}$ using Kalman smoothing \cite{kailath}.
When $\hat{\mathbf{c}}_i^{(l)}$ is given, from (\ref{eq:blocks}), the system equations for Kalman smoothing becomes
\begin{align}
\mathbf{s}_{n+1} &= \mathbf{A}_{n}\mathbf{s}_{n} + \mathbf{v}_{n} \nonumber \\ \label{eq:ssm}
\mathbf{y}_{n} &= \mathbf{B}_{n} {\rm diag}(\hat{\mathbf{c}}_i^{(l)})\mathbf{s}_{n} + \mathbf{w}_{n}.
\end{align}
%
%
We employ the fixed-interval Kalman smoothing algorithm performing sequential estimation of   $\widehat{\mathbf{s}}_{n|1:T}$ and  $\Sigma_{n|1:T}$ via forward and backward recursions in a block of observations $\mathbf{y}_{1:T}$.
 Let $\widehat{\mathbf{s}}_{n|j}$ and ${\Sigma}_{n|j}$ be the conditional mean and covariance given the first $j$ observation vectors, i.e.,   $\widehat{\mathbf{s}}_{n|j} = E\left[\mathbf{s}_{n}\bigg|  \mathbf{y}_{Ti+1}, \cdots, \mathbf{y}_{Ti+j} ; \hat{\mathbf{c}}_i^{(l)} \right]$ and  ${\Sigma}_{n|j} = {\rm Cov}\left[\mathbf{s}_{n}\bigg|  \mathbf{y}_{Ti+1}, \cdots, \mathbf{y}_{Ti+j} ; \hat{\mathbf{c}}_i^{(l)} \right]$, then the fixed-interval Kalman smoothing algorithm is summarized as
\begin{itemize}
\item \textbf{Forward recursion rule:}
\begin{align} \label{eq:f1}
\widehat{\mathbf{s}}_{n|n-1} &= \mathbf{A}_{n-1}\widehat{\mathbf{s}}_{n-1|n-1} \\
\Sigma_{n|n-1} &= \mathbf{A}_{n-1}\Sigma_{n-1|n-1}\mathbf{A}_{n-1}^{H}+\mathbf{V}_{n-1} \\
K_{n} &= \Sigma_{n|n-1}{\rm diag}(\hat{\mathbf{c}}_i^{(l)}) \mathbf{B}_{n}^{H}\left(\mathbf{B}_{n}{\rm diag}(\hat{\mathbf{c}}_i^{(l)})\Sigma_{n|n-1}{\rm diag}(\hat{\mathbf{c}}_i^{(l)})\mathbf{B}_{n}^{H}+\sigma_{w}^{2}\mathbf{I}\right)^{-1} \label{eq:kn} \\
\widehat{\mathbf{s}}_{n|n} &= \widehat{\mathbf{s}}_{n|n-1} + K_n(\mathbf{y}_n - \mathbf{B}_{n} {\rm diag}(\hat{\mathbf{c}}_i^{(l)}) \widehat{\mathbf{s}}_{n|n-1}) \\
\Sigma_{n|n}&= \left(I - K_{n}\mathbf{B}_{n} {\rm diag}(\hat{\mathbf{c}}_i^{(l)}) \right)\Sigma_{n|n-1}, \label{eq:fe}
\end{align}
\item \textbf{Backward recursion rule:}
\begin{align}\label{eq:b1}
\mathbf{S}_{n} &= \Sigma_{n|n}\mathbf{A}_{n}\Sigma_{n+1|n}^{-1} \\
\widehat{\mathbf{s}}_{n|1:T} &= \widehat{\mathbf{s}}_{n|n} + \mathbf{S}_n\left(\widehat{\mathbf{s}}_{n+1|1:T}-\mathbf{A}_{n}\widehat{\mathbf{s}}_{n|n}\right) \\
\Sigma_{n|1:T} &= \Sigma_{n|n} + \mathbf{S}_{n}\left(\Sigma_{n+1|1:T}-\Sigma_{n+1|n}\right)^{H}\mathbf{S}_{n}^{H}. \label{eq:be}
\end{align}
\end{itemize}
Using $\widehat{\mathbf{s}}_{n|1:T}$ and $\Sigma_{n|1:T}$, $Q\left(\mathbf{c}_{i}; \hat{\mathbf{c}}_i^{(l)}\right)$  can be rewritten as
\begin{align}
Q\left(\mathbf{c}_{i}; \hat{\mathbf{c}}_i^{(l)}\right)
=  & C'' + \frac{1}{\sigma_{w}^{2}} \sum_{n=Ti+1}^{T(i+1)} \bigg\{ 2 {\rm Re}\left(  \mathbf{y}_{n}^{H} \mathbf{B}_{n} {\rm diag}(\mathbf{c}_{i}) \widehat{\mathbf{s}}_{n|1:T} \right)  \nonumber \\
& \;\;\;\;\;\;\;\;\; - {\rm tr}\left[   \mathbf{B}_{n} {\rm diag}(\mathbf{c}_{i}) \left(\Sigma_{n|1:T} + \widehat{\mathbf{s}}_{n|1:T}\widehat{\mathbf{s}}_{n|1:T}^{H} \right)  {\rm diag}(\mathbf{c}_{i}) \mathbf{B}_{n}^{H} \right]  \bigg\}. \label{eq:qb}
\end{align}
Note that the second term in the right-hand side of (\ref{eq:qb}) is expressed as (see Appendix \ref{appen:prf})
\begin{align}
{\rm tr}&\left[   \mathbf{B}_{n} {\rm diag}(\mathbf{c}_{i}) \left(\Sigma_{n|1:T} + \widehat{\mathbf{s}}_{n|1:T}\widehat{\mathbf{s}}_{n|1:T}^{H} \right)  {\rm diag}(\mathbf{c}_{i}) \mathbf{B}_{n}^{H} \right] \nonumber \\
& \;\;\;\;\;\;\;\;\;\;\; = \mathbf{c}_{i}^{T} \left({\rm conj}\left(\mathbf{B}_{n}^{H} \mathbf{B}_{n}\right)  \odot     \left(\Sigma_{n|1:T} + \widehat{\mathbf{s}}_{n|1:T}\widehat{\mathbf{s}}_{n|1:T}^{H} \right) \right) \mathbf{c}_{i} \label{eq:sterm}.
\end{align}
From (\ref{eq:qb}) and (\ref{eq:sterm}), we have
\begin{align}
Q\left(\mathbf{c}_{i}; \hat{\mathbf{c}}_i^{(l)}\right)
=  & C''+  \frac{1}{\sigma_{w}^{2}}  \sum_{n=Ti+1}^{T(i+1)} \bigg\{ 2 {\rm Re}\left(  \mathbf{y}_{n}^{H} \mathbf{B}_{n} {\rm diag}(\widehat{\mathbf{s}}_{n|1:T})   \right) \mathbf{c}_{i}   \nonumber \\
& \;\;\;\;\;\;\;\;\;   - \mathbf{c}_{i}^{T} \left({\rm conj}\left(\mathbf{B}_{n}^{H} \mathbf{B}_{n}\right)  \odot     \left(\Sigma_{n|1:T} + \widehat{\mathbf{s}}_{n|1:T}\widehat{\mathbf{s}}_{n|1:T}^{H} \right) \right) \mathbf{c}_{i} \bigg\}. \label{eq:qb2}
\end{align}
Further, by denoting
\begin{align}
\mathbf{d}_i^{T} &= \sum_{n=Ti+1}^{T(i+1)} 2 {\rm Re}\left(  \mathbf{y}_{n}^{H} \mathbf{B}_{n} {\rm diag}(\widehat{\mathbf{s}}_{n|1:T})  \right)  \label{eq:cal1} \\
\mathbf{\Phi}_i &= \sum_{n=Ti+1}^{T(i+1)} \left({\rm conj}\left(\mathbf{B}_{n}^{H} \mathbf{B}_{n}\right)  \odot     \left(\Sigma_{n|1:T} + \widehat{\mathbf{s}}_{n|1:T}\widehat{\mathbf{s}}_{n|1:T}^{H} \right) \right),
\label{eq:cal2}
\end{align}
we have
\begin{align} \label{eq:finalexp}
Q\left(\mathbf{c}_{i}; \hat{\mathbf{c}}_i^{(l)}\right)
=  & C'' + \frac{1}{\sigma_{w}^{2}} \mathbf{d}_i^{T} \mathbf{c}_{i} - \frac{1}{\sigma_{w}^{2}} \mathbf{c}_{i}^{T}   \mathbf{\Phi}_i  \mathbf{c}_{i}.
\end{align}
In summary, the E-step performs the Kalman smoothing operation in (\ref{eq:f1})-(\ref{eq:be}) to estimate $\widehat{\mathbf{s}}_{n|1:T}$ and $\Sigma_{n|1:T}$ and also operations in  (\ref{eq:cal1}) and (\ref{eq:cal2}) to compute $\mathbf{d}_i^{T}$ and $\mathbf{\Phi}_i$ used in the computation of $Q\left(\mathbf{c}_{i}; \hat{\mathbf{c}}_i^{(l)}\right)$.

\subsection{\textbf{M-Step}}
\label{sec:mstep}
\begin{figure} [t]
 \centering
\centerline{\epsfig{figure=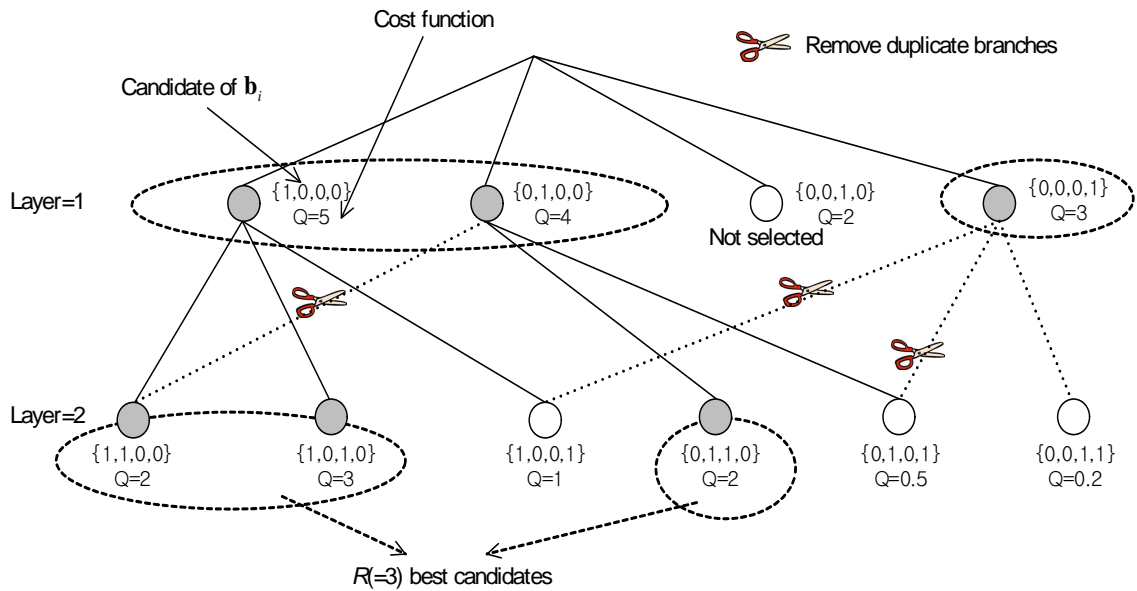, width=150mm}}
  \caption {Illustration of the proposed greedy tree search algorithm for the size of the source vector $M=4$ and the sparsity order $K=2$. Since the number of candidates chosen in each layer is set to $R=3$, $R(=3)$ nodes with the largest $Q$ values are chosen in each layer. }
  \label{fig:tree}
\end{figure}

\begin{table}[t]
\begin{center}
\caption{Summary of the greedy tree search algorithm}
\begin{tabular}{p{15cm}}
  \hline
  \hline
  Input: $\mathbf{d}_i$ and survival list $\mathbf{\Phi}_i$  \\ \hline
   Initialization:  Start with  $\Theta = \{[0, \cdots, 0]^{T}\}$. \\
  {\tt for} $k = 1:K$ \\
 \hspace{0.5cm} {\tt for} $j = 1:R$  \\
  \hspace{1cm} {\tt for} $m = 1:M$\\
   \hspace{1.5cm} Let ${\boldsymbol \theta}_j$ be the $j$th element of $\Theta$. \\
   \hspace{1.5cm} If the $m$-th entry is already one, skip the  loop. Otherwise, set the $m$-th entry of ${\boldsymbol \theta}_j$ to one. \\
   \hspace{1.5cm} For $Q(\mathbf{x}) \triangleq \mathbf{d}_i^{T} \mathbf{x} -\mathbf{x}^{T}   \mathbf{\Phi}_i \mathbf{x}$, evaluate $Q({\boldsymbol \theta}_j)-Q({\boldsymbol \theta})$ for all $ {\boldsymbol \theta} \in \Theta$. \\
   \hspace{1.5cm}  If $Q({\boldsymbol \theta}_j)-Q({\boldsymbol \theta})=0$ for any $ {\boldsymbol \theta} \in \Theta$, then the candidate ${\boldsymbol \theta}_j$ is duplicate node and hence we remove it. \\
  \hspace{1.5cm}   If  $Q({\boldsymbol \theta}_j)  > \min_{{\boldsymbol \theta} \in \Theta} Q({\boldsymbol \theta})$, add ${\boldsymbol \theta}_j$ into  $\Theta$. \\
     \hspace{1cm}  {\tt end}\\
    \hspace{0.5cm}  {\tt end}\\
        {\tt end}\\ \hline
              Output:   $\hat{\mathbf{c}}_i^{(l+1)} = \arg\max_{{\boldsymbol \theta} \in \Theta}Q( {\boldsymbol \theta})$.
              \\
   \hline
  \hline
  \end{tabular}
 \label{tb:tree}
  \end{center}
\end{table}

In the M-step, we find $\mathbf{c}_{i}$ maximizing $Q\left(\mathbf{c}_{i}; \hat{\mathbf{c}}_i^{(l)}\right)$  in (\ref{eq:finalexp}) as
\begin{align} \label{eq:max}
\hat{\mathbf{c}}_i^{(l+1)} = \arg \max_{\mathbf{c}_{i} \in \{0,1\}^M, \sum_{j=1}^{M}c_{i,j}=K}\widehat{Q}\left(\mathbf{c}_{i}; \hat{\mathbf{c}}_i^{(l)}\right)    ,
\end{align}
where $\widehat{Q}\left(\mathbf{c}_{i}; \hat{\mathbf{c}}_i^{(l)}\right) =\left(\mathbf{d}_i^{T} \mathbf{c}_{i} - \mathbf{c}_{i}^{T}   \mathbf{\Phi}_i  \mathbf{c}_{i}\right)$.
In finding $\hat{\mathbf{c}}_i^{(l+1)}$, we need to check all possible combinations satisfying the sparsity constraint $\sum_{j=1}^{M}c_{i,j}=K$. Since this brute force search is prohibitive for practical values of $M$, we consider a computationally efficient search algorithm returning a sub-optimal solution to the problem in (\ref{eq:max}).  The proposed approach, which in essence builds on the greedy tree search algorithm, examines candidate vectors to find out the most promising candidate of $\mathbf{c}_{i}$ in a cost effective manner.
The tree structure used for the proposed greedy search algorithm is illustrated in  Fig. \ref{fig:tree}.     Starting from a root node of the tree (associated with $\mathbf{c}_{i}=[0, \cdots, 0]^{T}$), we construct the layer of the tree one at each iteration. In the first layer of the tree, only one entry of $\mathbf{c}_{i}$ is set to one. For example, the nodes in the first layer of the tree are expressed as  $\mathbf{c}_{i}^{1} = [1, 0, \cdots, 0]^{T},\cdots, \mathbf{c}_{i}^{M} = [0, \cdots, 0, 1]^{T}$. As the layer increases, one additional entry is set to one and thus $K$ entries of $\mathbf{c}_{i}$ are set to one in the $K$-th layer ($\|\mathbf{c}_{i}\|_0=K$) (see Fig. \ref{fig:tree}). At each layer of the tree, we evaluate the cost function  $\widehat{Q}\left(\mathbf{c}_{i}; \hat{\mathbf{c}}_i^{(l)}\right)$  for each node and then choose the $R$ best nodes whose cost function is maximal. The rest of nodes are discarded from the tree. The candidates of $\mathbf{c}_{i}$ associated with the $R$ best nodes at each layer are called ``survival list".
For each node in the survival list, we construct the $M-1$ child nodes in the second layer by setting one additional entry of $\mathbf{c}_{i}$ to one\footnote{For example, if $\mathbf{c}_{i}= [1, 0, \cdots, 0]^{T}$ is in the survival list, then the child nodes of $\mathbf{c}_{i}$ becomes  $\mathbf{c}_{i} = [1, 1, \cdots, 0]^{T},\cdots,\mathbf{c}_{i} = [1, 0, \cdots, 1]^{T}$.}. Note that since we do not distinguish the order of the bit assertion in $\mathbf{c}_{i}$,  two or more nodes might represent the same realization of $\mathbf{c}_{i}$ during this process (see Fig. \ref{fig:tree}). When duplicate nodes are identified, we keep only one and discard the rest from the tree. After removing all duplicate nodes, we choose the $R$ best nodes  and then move on to the next layer. This process is repeated until the tree reaches the bottom layer of the tree. We note that since the tree search complexity is proportional to the depth of the tree ($K$), the dimension of source vector ($M$), and the number of nodes being selected ($R$), one can easily show that the complexity of the proposed tree search is $O(MRK)$. Hence, with small values of $R$ and $K$, the computational complexity is reasonably small and proportional to the dimension $M$ of the source signal vector. The proposed tree search algorithm is summarized in Table \ref{tb:tree}.

It is worth mentioning that one important issue to be considered is how to estimate the sparsity order $K$.
One simple way is to use the simple correlation method, where the observation vectors are correlated with the column vectors of $\mathbf{B}_{n}$ and $K$ is chosen as the number of the column vectors whose absolute correlation exceeds the predefined threshold.  While this approach is simple to implement, the performance might be affected by the estimation quality of $K$. One can alternatively consider a simple heuristic that terminates the tree search when a big drop in the cost metric $Q\left(\mathbf{c}_{i}, \hat{\mathbf{c}}_i^{(l)}\right)$ is observed.

After all iterations are finished (i.e. $l=L$) and $\hat{\mathbf{c}}_i^{(L)}$ is obtained, we use the Kalman smoother once again to compute $\widehat{\mathbf{s}}_{n|1:T}$ using the newly updated $\hat{\mathbf{c}}_i^{(L)}$.
The final estimate of $\mathbf{h}_{n}$  is expressed as
\begin{align}
\hat{\mathbf{h}}_{n} = {\rm diag}(\hat{\mathbf{c}}_i^{(L)}) \widehat{\mathbf{s}}_{n|1:T}. \label{eq:fnalest}
\end{align}

\begin{table*}[t]
\begin{center}
\caption{Summary of the proposed algorithm}
\begin{tabular}{p{15cm}}
  \hline
  \hline
Input:  $\mathbf{y}_{1:T}$, $\mathbf{B}_{1:T}$ for the $i$th block \vspace{-0.4cm}  \\ \hline
  STEP 1:  Set $l=0$. Start with $\hat{\mathbf{c}}_{i}^{(0)}= [1, \cdots, 1]^{T}$ and $K^{(0)}$.  \\
   STEP 2: Run Kalman smoother in (\ref{eq:f1}) to (\ref{eq:be}). \\
 STEP 3: Calculate $\mathbf{d}_{i}$ and $\mathbf{\Phi}_i$ in (\ref{eq:cal1}) and (\ref{eq:cal2}). \\
  STEP 4: Obtain $\hat{\mathbf{c}}_i^{(l+1)}$ by running the greedy tree search. \\
 STEP 5: Set $K \leftarrow K^{(l+1)}$, $l \leftarrow l+1$ and then go back to STEP 2. \\
 \hspace{1.2cm} If the estimate of $\mathbf{c}_{i}$ does not change (i.e., $\hat{\mathbf{c}}_i^{(l+1)}=\hat{\mathbf{c}}_i^{(l)}$) or the number of iterations reaches the limit, go to STEP 6. \\
 STEP 6: Run Kalman smoother using $\hat{\mathbf{c}}_{i}^{(L)}$. \\
   STEP 7: Obtain the final signal estimate from (\ref{eq:fnalest}). \\  \hline
              Output: $\widehat{\mathbf{h}}_{1:T}$
              \\
   \hline
  \hline
  \end{tabular}
 \label{tb:alg}
  \end{center}
\end{table*}

\subsection{Iteration Control}
\label{sec:sKTS_iter}

%
In this subsection, we discuss how to configure the control parameters in performing the iterations of the EM algorithm.
In each iteration, the proposed scheme estimates the  support $\mathbf{c}_{i}$ of the sparse signal vector under the sparsity  constraint $\sum_{j=1}^{M}c_{i,j}=K$.
Since the tree search to identify the support of $\mathbf{h}_{n}$ is based on the greedy principle, it is possible that the support elements might not be accurately identified,  especially for the initial iterations where the cost metric $\widehat{Q}\left(\mathbf{c}_{i}; \hat{\mathbf{c}}_i^{(l)}\right)$ is not so accurate.
In order to reduce the chance of missing nonzero entries of sparse vector in early iterations, we search for the sparse signal vector $\mathbf{c}_{i}$ under relaxed sparsity constraint in the beginning and then gradually reduce the sparsity order $K$ as the number of iterations increases. Let the sparsity order parameter used for the $l$th iteration be $K^{(l)}$. Then, we use sufficiently large value of $K^{(1)}$ initially\footnote{Initially, we set $\mathbf{c}_{i}^{(0)} = [1, \cdots, 1]$ (i.e., $K^{(0)}=M$) since  we have no knowledge on the sparsity structure of the source signal vector in the beginning.} and then decreases $K^{(l)}$ monotonically (i.e., $K^{(l)} \geq K^{(l+1)}$) until $K^{(l)}$ equals the target sparsity order $K$.
In doing so, we can substantially reduce the chance of missing support elements and at the same time gradually improve the estimation quality of $\mathbf{c}_{i}$.
 %
%
  %
%
The summary of the proposed algorithm is presented in Table \ref{tb:alg}.

\subsection{Estimation of $\mathbf{V}_{n}$}

Although the proposed algorithm is designed based on the dynamic sparse model in (\ref{eq:blocks}), $\mathbf{A}_{n}$ and $\mathbf{V}_{n}$ are generally unknown in practice. We are often interested in the scenario where the elements of the signal vector $\mathbf{s}_{n}$ are uncorrelated with each other and $\mathbf{A}_{n}$ and $\mathbf{V}_{n}$ are fixed over the block of interval $T$. Then, $\mathbf{A}_{n}$ and $\mathbf{V}_{n}$ have a diagonal form, i.e.,  $\mathbf{A}_{n}={\rm diag}(\alpha_1, \cdots, \alpha_{M})$ and $\mathbf{V}_{n}={\rm diag}((1-\alpha_1^2)\sigma_{s,1}^2, \cdots, (1-\alpha_M^2)\sigma_{s,M}^2)$. In general, $\alpha_j$ can be determined using the temporal correlations of the $j$th element of $\mathbf{s}_{n}$, which can be known a priori or estimated from the data separately.
While the estimation of $\alpha_i$ is relatively easy, such is not the case for $\sigma_{s,j}^2$. One accurate but computationally expensive approach is to estimate these variances using EM formulation. In this work, we do not pursue this approach due to complexity concern and compute a rough estimate of $\sigma_{s,j}^2$ from $\frac{1}{T}\sum_{n=Ti+1}^{T(i+1)}\mathbf{b}_{n,j}^{H}\mathbf{y}_{n}/\|\mathbf{b}_{n,j}\|_2^2$, where $\mathbf{b}_{n,j}$ is the $j$th column of the matrix $\mathbf{B}_{n}$. After running all iterations, the refined estimate of $\mathbf{V}_{n}$ can be obtained by taking sample covariance matrix of the estimated signal vectors in the processing block and used for the final Kalman smoothing step described in Section \ref{sec:estep}.

\subsection{Real-time Implementation}
\label{sec:sKTS_online}

\begin{figure} [t]
 \centering
\centerline{\epsfig{figure=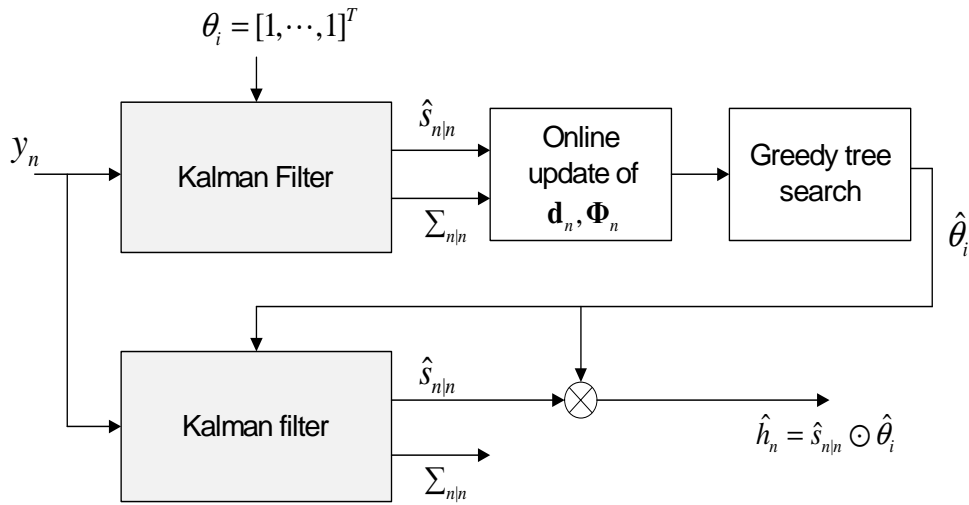, width=130mm}}
  \caption{Real-time implementation of the sKTS algorithm.}
  \label{fig:online}
\end{figure}

Since the proposed algorithm we described in the previous subsections performs batch processing by running several iterations of E-step and M-step in a block, it might not be suitable for real-time applications.
By slightly modifying the algorithm, we can reduce latency and also speed up the operations substantially. The main idea behind this modification is to return the estimate of the source signal immediately after the new measurement vector is provided.
In doing so, we can process the block seamlessly without waiting for the reception of whole block of observations.
First, instead of Kalman smoother, we employ the Kalman filter to conduct the operations from (\ref{eq:f1}) to (\ref{eq:fe}) in a forward direction. In order to ensure the real-time processing, we need to use multiple Kalman filters, where each Kalman filter corresponds to single iteration of the EM algorithm.
For the sake of simplicity, we here consider two Kalman filters as an example\footnote{This setup corresponds to single EM iteration.}. In the first Kalman filter, we do not know the signal existence vector $\mathbf{c}_{i}$ so that we set $\mathbf{c}_{i}=[1, \cdots, 1]^{T}$ and run the Kalman filter.  Once $\widehat{\mathbf{s}}_{n|n}$ and $ \Sigma_{n|n}$ are obtained, we replace the computation of $\mathbf{d}_{n}^{H}$ and $\mathbf{\Phi}_{n}$ in (\ref{eq:cal1}) and (\ref{eq:cal2}) by an  auto-regressive update rule
%
\begin{align}
\widetilde{\mathbf{d}}_{n}^{H} &= (1-\alpha) \widetilde{\mathbf{d}}_{n-1}^{H}  +  \alpha 2 {\rm Re}\left(  \mathbf{y}_{n-1}^{H} \mathbf{B}_{n-1} {\rm diag}(\widehat{\mathbf{s}}_{n-1|n-1})  \right) \label{eq:rup1} \\
\widetilde{\mathbf{\Phi}}_{n} &= (1-\alpha) \widetilde{\mathbf{\Phi}}_{n-1} + \alpha \left({\rm conj}\left(\mathbf{B}_{n-1}^{H} \mathbf{B}_{n-1}\right)  \odot     \left(\Sigma_{n-1|n-1} + \widehat{\mathbf{s}}_{n-1|n-1}\widehat{\mathbf{s}}_{n-1|n-1}^{H} \right) \right), \label{eq:rup2}
\end{align}
where $\alpha$ is a forgetting factor controlling the speed of update. 
Note that by using (\ref{eq:rup1}) and (\ref{eq:rup2}) instead of  (\ref{eq:cal1}) and (\ref{eq:cal2}),  we can compute approximation of $\mathbf{d}_{n}^{H}$ and $\mathbf{\Phi}_{n}$ on the fly whenever the new measurement vector is available.
Once $\widetilde{\mathbf{d}}_{n}^{H}$ and $\widetilde{\mathbf{\Phi}}_{n}$ are obtained, we next identify the signal existence indication vector $\mathbf{c}_{i}$ using the greedy tree search described in Section \ref{sec:mstep}.
Using the newly obtained estimate  $\hat{\mathbf{c}}_{i}$,  the second Kalman filter generates $\widehat{\mathbf{s}}_{n|n}$. By multiplying this and $\hat{\mathbf{c}}_i$, we get the final estimate of $\mathbf{h}_{n}$.
%
%
Note that it is straightforward to employ more than two Kalman filters to produce better estimate of $\mathbf{c}_{i}$. To distinguish this from the original sKTS algorithm,  in the sequel, we refer it to as real-time sKTS (RT-sKTS) algorithm.
The block diagram of the RT-sKTS algorithm is depicted in Fig. \ref{fig:online}.

%

\subsection{Convergence Behavior}
Since the proposed sKTS algorithm is derived based on EM algorithm, we can study the convergence  of the EM algorithm to understand the  behavior of the sKTS algorithm. In general, the convergence of the EM algorithm can be shown in two steps. First step is to show that the likelihood function $\Pr\left(\mathbf{y}_{1:T}; \mathbf{c}_{i}^{(l)}\right)$ is non-decreasing function of the iteration. Using the standard analysis for EM algorithm, one can show that the sKTS algorithm satisfies this property.
The second step is to show that $\mathbf{c}_{i}^{(l)}$ converges to a stationary point of the likelihood function $\Pr\left(\mathbf{y}_{1:T}; \mathbf{c}_{i}\right)$. Since we do not use the continuous model for $\mathbf{c}_{i}^{(l)}$ (the parameter vector $\mathbf{c}_{i}^{(l)}$ is binary), unfortunately, it is not easy to prove this convergence property. Nevertheless, we show from numerical experiments that the sKTS algorithm finds an accurate estimate of $\mathbf{c}_{i}$ in a small number of iterations.

\section{Application to Wireless Channel Estimation}

\label{sec:chan}

In this section, we study the application of the proposed scheme to the training-based channel estimation problem in wireless communication systems. In many communication systems, estimation of channels is done before the symbol detection since the channel estimate is required for the detection of the transmitted symbols. Also, to perform the precoding and user scheduling in the transmitter, accurate estimate of the channel vector should be fed back from the receiver to the transmitter. Since the wireless channels whose delay spread is larger than the number of significant paths are well modeled as a sparse signal vector in a discretized delay domain, the CS techniques have been used in the sparse channel estimation problem  \cite{bajwa,berger,mp_cs}. While existing approaches perform the sparse channel estimation using only a single observation vector or multiple observation vectors under the assumption that the CIR vector is invariant in the block, the proposed sKTS algorithm exploits the simultaneously sparse structure of time-domain CIR, which matches well with physical characteristics of multi-path fading channels.  In this section, we describe the application of the proposed method to the channel estimation problem in the orthogonal frequency division multiplexing (OFDM) and the single carrier (SC) systems.


%
\subsection{OFDM systems}
\label{sec:chan_ofdm}
%

\begin{figure} [!]
 \centering
    \subfigure[]
  {\epsfig{figure=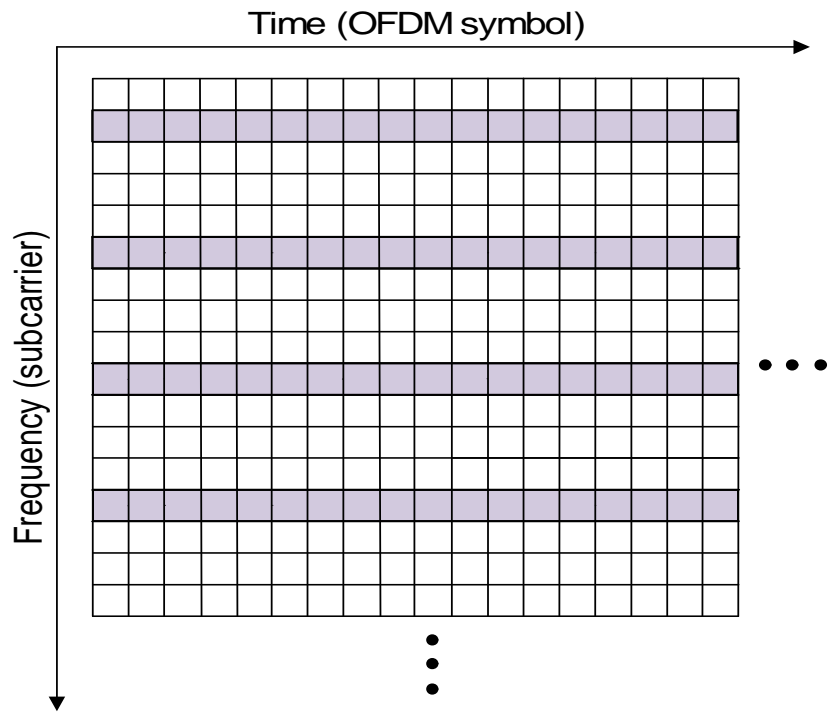, height = 95mm, width=110mm}}
   \subfigure[]
  {\epsfig{figure=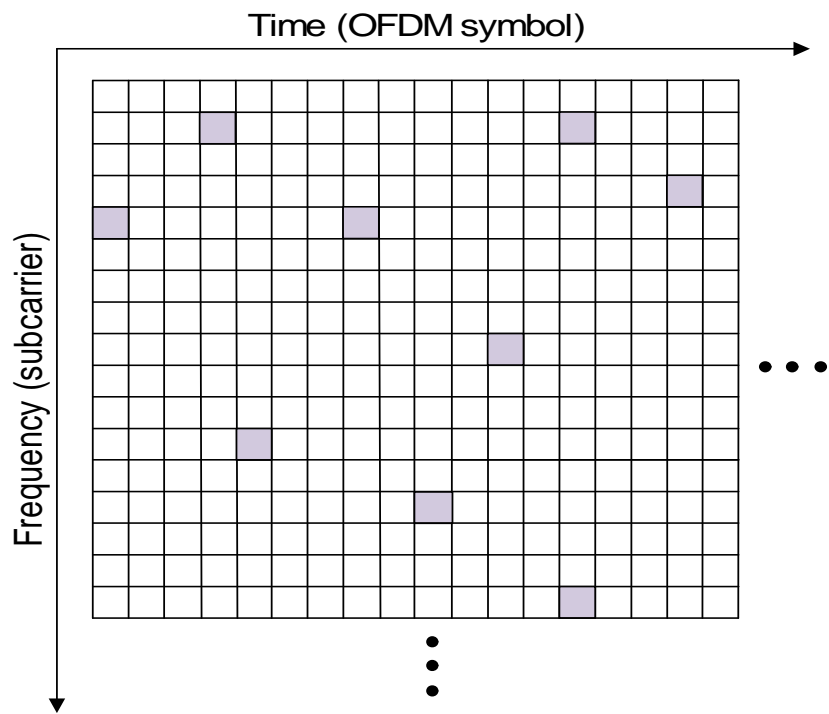, height = 95mm, width=110mm}}
  \caption { Pilot allocation in (a) conventional systems (comb-type assignment) vs. (b) proposed sKTS algorithm.
   } \label{fig:pilot}
\end{figure}

We first consider the channel estimation problem of the OFDM systems.
In our simulations, we focus on the scenario where the number of the pilot symbols transmitted per OFDM symbol is much smaller than the length of the CIR, thereby forming underdetermined systems in the estimation of the CIR.  Note that this scenario will be prevalent when a  large number of transmit antennas are deployed (e.g., in large-scale multi-input multi-output systems) since the required number of the pilot signals is proportional to the number of the transmit antennas. Since too much pilot overhead will eat out the resources and eventually limit the throughput of the systems, it is desirable to estimate the channel with small number of resources.
However, when the number of the pilot signals is small,  conventional channel estimators do not perform well due to the lack of observations. Whereas, by exploiting the sparsity structure of the CIR vector, the sKTS algorithm overcomes the shortage of pilot signals. In the proposed scheme, as shown in  Fig.~\ref{fig:pilot}, we randomly allocate the pilot signals in time and frequency axis to make the better conditioned system matrix $\mathbf{B}_n$. As a result, while the support of $\mathbf{h}_{n}$ is invariant for several OFDM symbols, the composite system matrix is varying per symbol.

 Let $P$, $N$, and $M$ be the total number of the subcarriers, the number of the pilot subcarriers, and the length of the time-domain CIR, respectively. Further, let $N$ be the number of pilot subcarriers per OFDM symbol. Then the relationship between the pilot signal vector $\mathbf{p}_{n} \in {\mathcal C}^{N}$ and the observed signal vector $\mathbf{y}_{n} \in {\mathcal C}^{N}$ of the OFDM system is expressed as
\begin{align}
\mathbf{y}_{n} =  \mathbf{p}_{n} \odot  \mathbf{g}_{n} + \mathbf{v}_{n} \\
 =  {\rm diag}(\mathbf{p}_{n})  \mathbf{g}_{n} + \mathbf{v}_{n} \label{eq:ofdm}
\end{align}
where  $\mathbf{g}_{n}$ is the vector representing frequency-domain channel response.
Using the $P \times P$ DFT matrix $F_{P}$ whose $(i,j)$ entry is
 given by $e^{-\frac{j2\pi}{P}ij}$, the frequency-domain channel response is expressed in terms of the time-domain CIR as
\begin{align}
 \mathbf{g}_{n} = \Pi_{n} F_{P} \Phi \mathbf{h}_{n}, \label{eq:ofdm2}
\end{align}
where $\mathbf{h}_{n}$ is the $M\times 1$ vector representing the time-domain CIR and   $\Pi_{n}$ is the $N \times P$ matrix that selects the $N$ rows of $F_{P}$ depending on the location of pilot subcarriers, i.e.,
\begin{align}
  \Pi_{n} = \begin{bmatrix}\mathbf{e}_{c_{n,1}}^{T} \\ \vdots \\  \mathbf{e}_{c_{n,N}}^{T}\end{bmatrix}.
\end{align}
Recall that  $\{c_{n,1}, c_{n,2}, \cdots, c_{n,N}\}$ are the  pilot subcarrier indices at the $n$th OFDM symbol and $\Phi$ is the $P\times M$ matrix choosing the first $M$ columns of $F_{P}$, i.e.,
\begin{align}
 \Phi = \begin{bmatrix}\mathbf{e}_{1} & \cdots &  \mathbf{e}_{M}\end{bmatrix}.
\end{align}
Using (\ref{eq:ofdm}) and (\ref{eq:ofdm2}), we obtain the observation model for which the proposed sKTS scheme can be readily applied
\begin{align}
 \mathbf{y}_{n}  =   \mathbf{B}_{n} \mathbf{h}_{n}   + \mathbf{w}_{n},   \label{eq:ofdmfor2}
\end{align}
where $\mathbf{B}_{n} = {\rm diag}(\mathbf{p}_{n}) \Pi_{n} F_{P} \Phi$.

%
\subsection{Single Carrier Systems}
\label{sec:chan_sc}
%

In the single carrier (SC) transmission system, the known training symbols are sent from the transmitter to the receiver before the transmission of the data symbols.
%
Suppose that the length of the training symbols being transmitted is $N$, then the  received  signal at time $n$  is expressed as
\begin{align} \label{eq:train_ob}
y_n = \sum_{l=1}^{M} h_{n,l}t_{n-l+1} + w_{n}, \;\;\;\; 0  \leq n \leq N,
\end{align}
where $h_{n,l}$ is the $l$th tap of the CIR at time $n$ and $t_{i}$ is the $i$th training symbol.
\begin{table}[!]
\caption{Simplified Algorithm  For Channel Estimation in SC Systems}
\begin{center}
\begin{tabular}{l|c}
\hline
 Operational step & Algorithm change \\ \hline  \hline
Eq.  (\ref{eq:kn}) & $K_{n} = \frac{\Sigma_{n|n-1}{\rm diag}(\hat{\mathbf{c}}_i^{(l)}) \mathbf{t}_{n}}{\mathbf{t}_{n}^{H}{\rm diag}(\hat{\mathbf{c}}_i^{(l)})\Sigma_{n|n-1}{\rm diag}(\hat{\mathbf{c}}_i^{(l)})\mathbf{t}_{n}+\sigma_{w}^{2}} $ \\ \hline
  Eq. (\ref{eq:cal1}) & $\mathbf{d}_i^{T} \triangleq \sum_{n=Ti+1}^{T(i+1)} 2 {\rm Re}\left(  conj(y_{n}) \mathbf{B}_{n} {\rm diag}(\widehat{\mathbf{s}}_{n|1:T})  \right)$ \\  \hline
  Eq. (\ref{eq:cal2}) & $\mathbf{\Phi}_i \triangleq \sum_{n=Ti+1}^{T(i+1)} {\rm diag}(\mathbf{t}_{n}^{H})     \left(\Sigma_{n|1:T} + \widehat{\mathbf{s}}_{n|1:T}\widehat{\mathbf{s}}_{n|1:T}^{H} \right) {\rm diag}(\mathbf{t}_{n})$ \\  \hline
  \end{tabular}
\end{center}
\label{tb:eq}
\end{table}
%
%
Using a vector notation, we have
\begin{align}
y_{n}&=\mathbf{t}_{n}^{H} \mathbf{h}_n + w_{n},  \;\;\;\; 0  \leq n \leq N, \label{eq:obs}
\end{align}
where $\mathbf{h}_{n} = [h_{n,1}, \cdots, h_{n,M}]^{T}$ and $\mathbf{t}_{n} = [t_{n}^{*}, \cdots, t_{n-M-1}^{*}]^{T}$.
To estimate the channel vector $\mathbf{h}_n$, recursive least square (RLS) or Kalman channel estimators have been popularly used \cite{akino, romano}.
When the training period $N$ is small, the system becomes ill-posed and thus the estimation quality of these channel estimation algorithms would be severely degraded.
Whereas, by exploiting the simultaneously sparse structure of $\mathbf{h}_{n}$ and treating $\mathbf{t}_{n}$ as a system matrix, the sKTS algorithm generates reliable estimate of $\mathbf{h}_{n}$.
Note that since the system matrix is a row vector $\mathbf{t}_{n}^{H}$ in (\ref{eq:obs}), each step of the algorithm can be simplified (see Table \ref{tb:eq}).

%
\section{Simulations and Discussion}
\label{sec:simul}
%

In this section, we study the performance of the proposed sKTS algorithm.
We first conduct simulations with the synthetic data, and then we test the performance of sKTS algorithm in wireless channel estimation and MRI image reconstruction problem.
In our simulations, we compare the performance of the following algorithms:
\begin{itemize}
\item Proposed sKTS algorithm: we set the tree search parameter $R$ to $5$. Only two iterations with $K^{(0)}=2K$ and $K^{(1)}=K$ are performed.
\item Conventional Kalman smoother: standard Kalman smoother \cite{kailath} is used.
\item Oracle-based Kalman smoother: Kalman smoothing is performed under the perfect knowledge on the support of the CIR.  This algorithm provides the best achievable performance bound of the proposed sKTS algorithm.
\item OMP algorithm \cite{mp_cs}: greedy projection is used to estimate the signal support.
\item KSBL algorithm \cite{prasad}: the SBL algorithm \cite{rao_temporal} is extended to sequential signal estimation based on the autoregressive model.
\item RW1L-DF algorithm \cite{charles2}: the reweighted $\ell_1$ optimization is modified to perform sequential dynamic filtering.
\end{itemize}
As a metric to measure the estimation performance, we use the normalized mean square error (MSE) defined as
\begin{align}
{\rm MSE} = 10\log_{10}\frac{E[\|\mathbf{h}_{n}-\widehat{\mathbf{h}}_{n}\|^{2}]}{E[\|\mathbf{h}_{n}\|^2]}.
\end{align}

%
\subsection{Experiments with synthetic data}
%

%
\subsubsection{Simulation Setup}
%
We first evaluate the performance of the sKTS algorithm using the synthetic data generated from the signal model in (\ref{eq:ssm}). Here we assume that $\mathbf{A}_{n}$ and $\mathbf{V}_{n}$ are given by $\alpha \mathbf{I}_{M \times M}$ and $\sigma_{s}^2 (1 - \alpha^2) \mathbf{I}_{M \times M}$. The location of nonzero elements of $\mathbf{h}_{n}$ is fixed over the interval of $T=30$ and changes randomly for different intervals.
Note that slow variation of support is not considered in the simulations. The signal dimension $M$ is set to $200$ and the sparsity order $K$ is set to $15$. Entries of the matrix $\mathbf{B}_{n}$ are generated from i.i.d. Gaussian distribution $\mathcal{N}(0,1/M)$.
%

%
\subsubsection{Simulation Results}
\begin{figure} [!]
 \centering
   \subfigure[]
  {\epsfig{figure=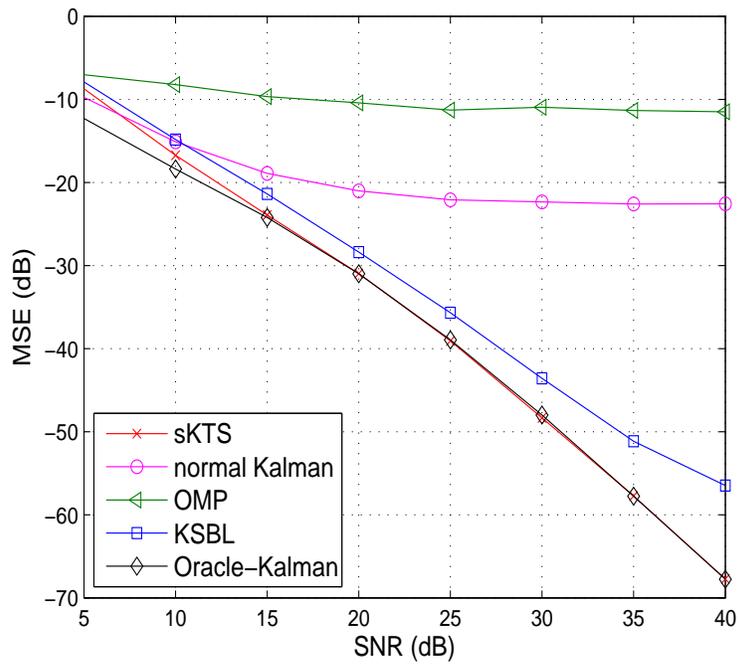, height = 95mm, width=110mm}}
   \subfigure[]
  {\epsfig{figure=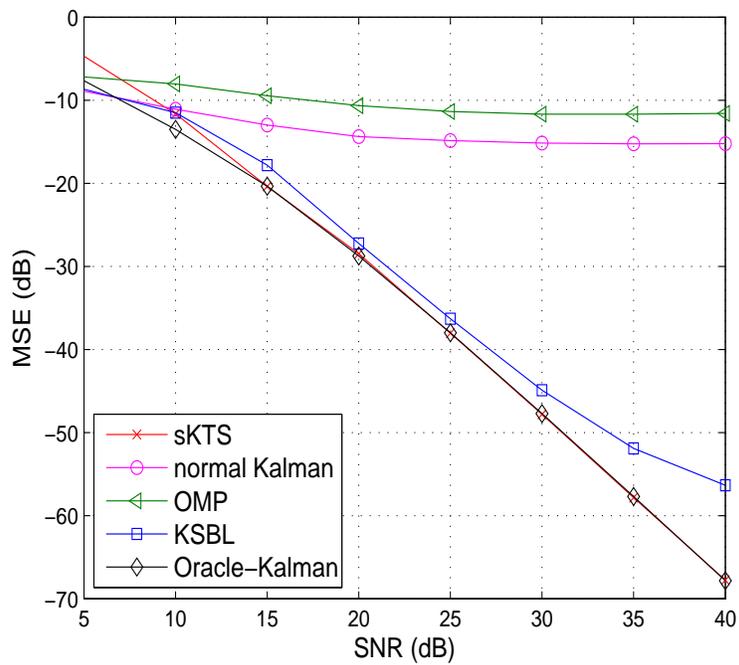, height = 95mm, width=110mm}}
	   \caption { The MSE performance as a function of SNR with (a) $\alpha = 0.8$ and (b) $\alpha=0$.
   } \label{fig:alpha}
\end{figure}

%
Fig.~\ref{fig:alpha} presents the MSE performance of signal recovery algorithms as a function of the SNR. In Fig.~\ref{fig:alpha} (a) and (b), the auto-regressive parameter $\alpha$ is set to $0.8$ and $0$, respectively.
Note that when $\alpha=0$, each element of $\mathbf{s}_{n}$ is temporally uncorrelated. For both scenarios, the dimension $N$ of the measurement vector is set to $40$. As shown in the figure, the proposed sKTS algorithm outperforms competing recovery algorithms and also performs close to the Oracle-based Kalman smoother for the whole SNR regime.
Since the dimension $N$ of the measurement vector is much smaller than the dimension $M$ of the source signal ($N \ll M$), it is no wonder that the conventional Kalman smoother does not perform well. Since the OMP algorithm uses each measurement vector independently, its performance is also not appealing.

\begin{figure} [!]
 \centering
\centerline{\epsfig{figure=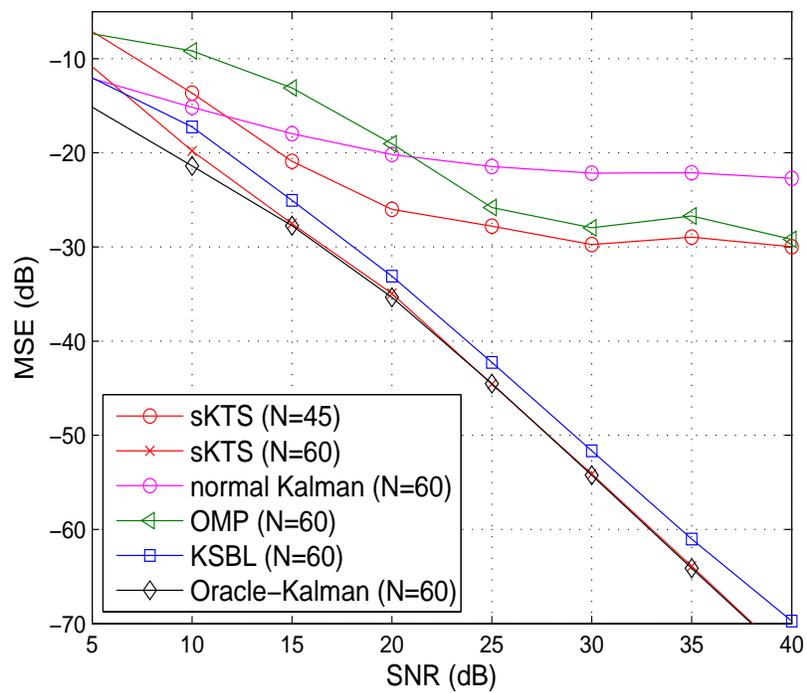, height=100mm, width=120mm}}
  \caption {The MSE performance when $\mathbf{B}_{n}$ is fixed.}
  \label{fig:bnfix}
\end{figure}

Next, we investigate the MSE performance when the system matrix $\mathbf{B}_{n}$ does not change over the processing  block.  Since the system matrix is fixed, if $\mathbf{B}_{n}$ is ill-conditioned, recovery algorithm suffer from severe performance loss.  In order to alleviate this phenomenon, we increase the dimension $N$ of the measurement vector to $60$ and also use the iteration control by setting $K^{(0)}=4K$, $K^{(1)}=2K$, and $K^{(2)}=K$ (that is, we perform three iterations with different sparsity parameters). Fig. \ref{fig:bnfix} shows the MSE performance as a function of SNR. While the sKTS algorithm suffers from considerable performance loss when $N$ is set to $45$, it performs well when $N$ becomes $65$.

\begin{figure} [!]
 \centering
\centerline{\epsfig{figure=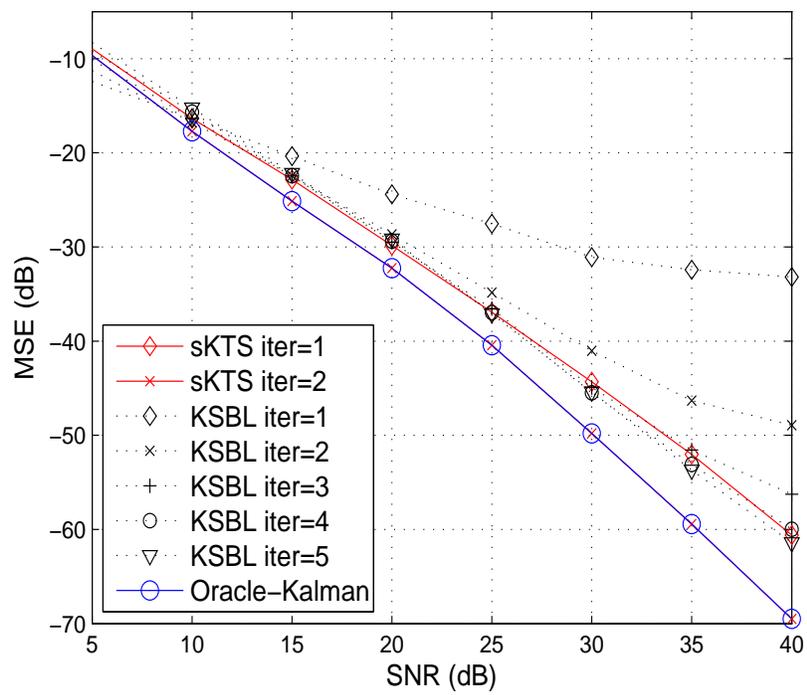, height=100mm, width=120mm}}
  \caption {The MSE performance of the sKTS and KSBL as a function of the number of iterations.}
  \label{fig:iter}
\end{figure}

We next take look at the convergence behavior of the sKTS algorithm and the KSBL algorithm. In this test, we set $N$ and $\alpha$ to 45 and 0.8, respectively. As shown in Fig.~\ref{fig:iter}, the sKTS algorithm performs close to the Oracle-based Kalman smoother after two iterations, while the KSBL algorithm requires five iterations until the performance converges.
Due to this reason, even though the computational complexity of the greedy tree search is a bit higher than the complexity of M-step in the EM algorithm, overall complexity of the approaches are more or less similar. In fact, our numerical experiments demonstrate that it takes  15.69 seconds for the sKTS algorithm to finish recovery process while the time required for the KSBL is 14.97 seconds.

%
\subsection{Experiments in application to channel estimation}
%
In this subsection, we study the performance of the proposed scheme in application to channel estimation in OFDM systems.

%
\subsubsection{Simulation Setup}
%
\label{sec:simul_setup}
 \begin{table}[!]
\caption{Parameters of the OFDM systems}
\begin{center}
\begin{tabular}{l|c}
\hline
 Setup & Specification  \\ \hline  \hline
Total number of subcarriers & $1024$ \\ \hline
  Bandwidth of each subcarrier & $15$ kHz \\  \hline
  Symbol duration & $66.7$ $\mu$s \\  \hline
  CP length & $16.7$ $\mu$s \\  \hline
   Interval between two consecutive pilot signals & $0.25$ $m$s (three OFDM symbol) \\ \hline
  Maximum delay spread of CIR & $13$  $\mu$s \\ \hline
  Modulation order & QPSK \\ \hline
  Code rate & 1/2 \\ \hline
  \end{tabular}
\end{center}
\label{tb:setup}
\end{table}
 The specific parameters for the OFDM system are summarized in the Table \ref{tb:setup}.
In generating pilot symbols, we use the quadrature phase shift keying (QPSK) pseudo-random sequence. The pilot signals are transmitted for every three OFDM symbols. For the OFDM symbol containing pilots, we assign $N$ pilot symbols. As described in the previous section, the location of pilot subcarriers is randomly chosen for each OFDM symbol. The remaining OFDM resources are filled with data symbols. The binary information bits are encoded using half rate convolutional code with generation polynomials $(171,133)$ and the coded bits are modulated
to  QPSK symbols. Each code block contains 23,360 coded bits.
 Considering the maximum channel delay spread specified in the Table \ref{tb:setup}, we set the dimension $M$ of  $\mathbf{h}_{n}$ to $200$.
In generating the complex Rayleigh fading frequency-selective channels, we use Jake's model \cite{jake}, where  temporal correlation of the CIR taps for given Doppler frequency $f_d$ (Hz) is expressed as $J_{0}(2\pi f_d T_s )$, where $T_s$ is the interval between consecutive pilot symbols in time ($T_s=0.25m$s) and $J_0(x) = \sum_{m=0}^{\infty} \frac{(-1)^{m}}{m! \Gamma(m+1)} \left( \frac{x}{2} \right)^{2m}$ is the zero-th order Bessel function.  For convenience, we use \emph{Doppler rate} defined as Doppler frequency normalized by pilot transmission rate, i.e.,  $f_dT_s$.
We use two types of channel models: 1) the exact $K$-sparse channel model where the location of the $K$ nonzero taps is randomly chosen for every block of $T$ OFDM symbols and 2) the practical channel models specified by 3GPP Long Term Evolution (LTE) standard \cite{lte101} (see  Table~\ref{tb:lte}).
%
 \begin{table}[!]
\caption{3GPP channel models}
\begin{center}
\begin{tabular}{l|l}
\hline
Extended Pedestrian A model (EPA)  & Delay = $[0,30,70,90,110,190,410]$ $n$s  \\ \cline{2-2}
 & Power =   $[0, -1, -2, -3, -8, -17.2, -20.8]$ dB\\ \hline
 Extended Vehicular A model  (EVA) & Delay = $[0,30,150,310,370,710,1090,1730,2510]$ $n$s  \\ \cline{2-2}
 & Power =   $[0, -1.5, -1.4, -3.6, -0.6, -9.1, -7.0, -12.0, -16.9]$ dB\\ \hline
Extended Typical Urban model (ETU) & Delay = $[0,50,120,200,230,500,1600,2300,5000]$ $n$s  \\ \cline{2-2}
 & Power =   $[-1, -1, -1, 0, 0, 0, -3,-5,-7]$ dB\\ \hline
  \end{tabular}
\end{center}
\label{tb:lte}
\end{table}
Note that the channel taps in the standard LTE channel model are only approximately sparse.
In order to determine the parameters of the Gauss-Markov process $\mathbf{A}_{n}$ and $\mathbf{V}_{n}$  for a given $f_d$, we minimize the approximation error between  the Gauss-Markov process and the Jake's model as suggested in \cite{baddour}. Using the CIR estimates obtained by the sparse signal recovery algorithms, the transmitted symbols are detected by the MMSE equalizer in frequency domain. Then, the channel decoder is followed to detect the information bits.
To evaluate the performance of the recovery algorithms, we measure bit error rate (BER) at the output of the channel decoder.



%
\subsubsection{Simulation Results}
\label{sec:simul_result}
%

\begin{figure} [!]
 \centering
  \subfigure[]
  {\epsfig{figure=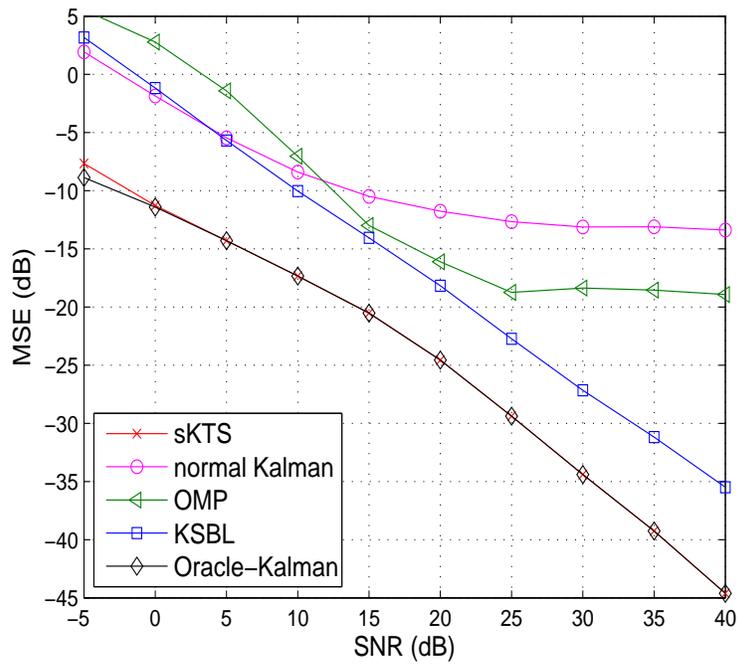, height = 95mm, width=110mm}}
   \subfigure[]
  {\epsfig{figure=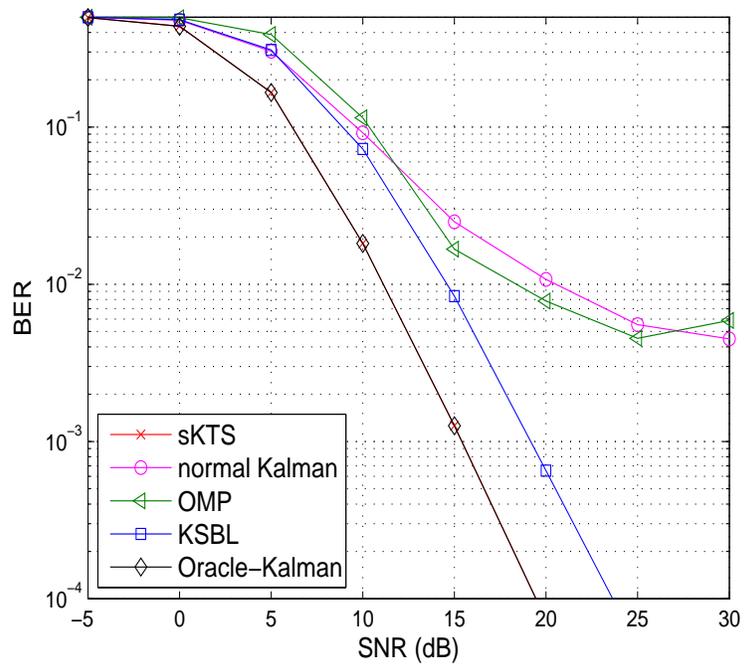, height = 95mm, width=110mm}}
	   \caption { The performance of the recovery algorithms for the $K$-sparse channel model: a) MSE and b) BER performance.
   } \label{fig:perf}
\end{figure}

We test the performance of the channel estimators when the exact $K$-sparse channels are used. The sparsity order $K$ for these channels is set to $8$ and the dimension $N$ of the measurement vector is set to $32$. Note that when $N=32$, the pilot resources occupy 3.12\%  of  the overall OFDM resources. We assume that the sparsity structure remains unchanged over the block of $T=30$ pilot containing OFDM symbols. We set the Doppler rate $D_r$  to $0.05$. In Fig. \ref{fig:perf} (a) and (b),  we plot the MSE and BER performance of the recovery algorithms as a function of SNR.
From the figure, we clearly observe that the sKTS algorithm performs best among all algorithms under test and also performs close to that of the Oracle-based Kalman smoother.

\begin{figure} [!]
 \centering
  \subfigure[]
  {\epsfig{figure=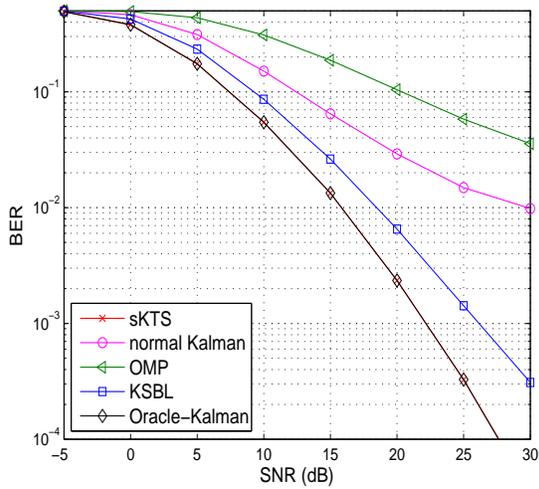, height = 70mm, width=80mm}}
   \subfigure[]
  {\epsfig{figure=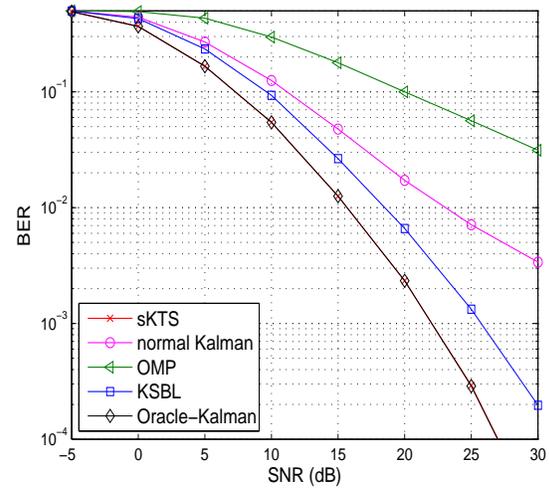, height = 70mm, width=80mm}}
   \subfigure[]
  {\epsfig{figure=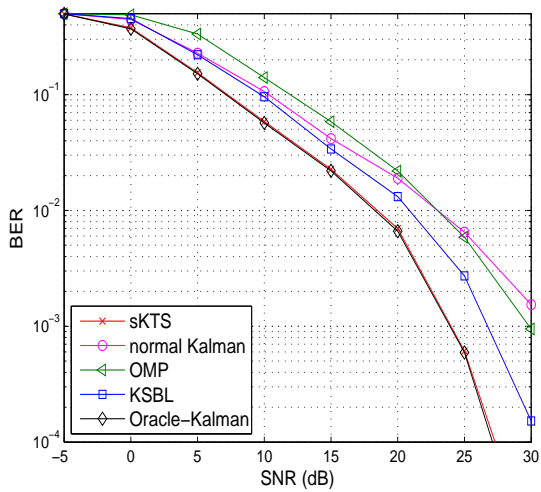, height = 70mm, width=80mm}}
   \subfigure[]
  {\epsfig{figure=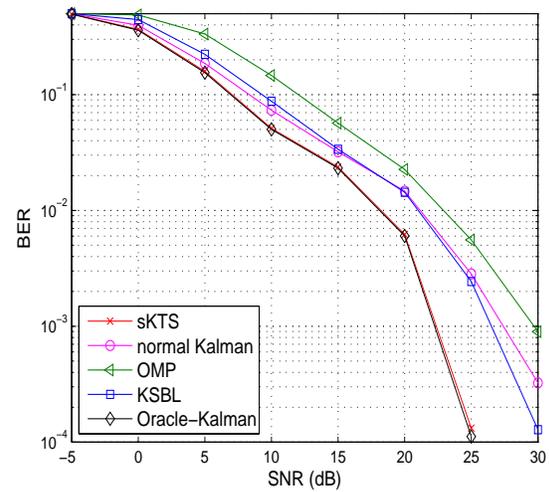, height = 70mm, width=80mm}}
	   \caption { The BER performance as a function of SNR for the four distnictive scenarios of a) EVA channel with $D_r = 0.1$ and $N =60$, b) EVA channel with $D_r = 0.05$ and $N =60$, c) EPA channel with $D_r = 0.02$ and $N =48$, and d) EPA channel with $D_r = 0.005$ and $N =48$.
   } \label{fig:3gpp}
\end{figure}

We next investigate the performance of the proposed sKTS algorithm when the practical LTE channel models are used. In this test, we observe the behavior of the algorithms for  four distinctive scenarios: a) EVA channel with $D_r = 0.1$ and $N =60$, b) EVA channel with $D_r = 0.05$ and $N =60$, c) EPA channel with $D_r = 0.02$ and $N =48$, and d) EPA channel with $D_r = 0.005$, $N =48$. We set $K=60$ and $K=24$ for EVA and EPA channel models since the EVA channel exhibits longer delay spread. In Fig.~\ref{fig:3gpp}, we observe that the sKTS algorithm maintains the performance gain over the competing algorithms for wide range of Doppler rates. Note that when compared to the results of the exact $K$-sparse channel model, we see that the performance gap between the sKTS and KSBL is a bit reduced.

\begin{figure} [!]
 \centering
\centerline{\epsfig{figure=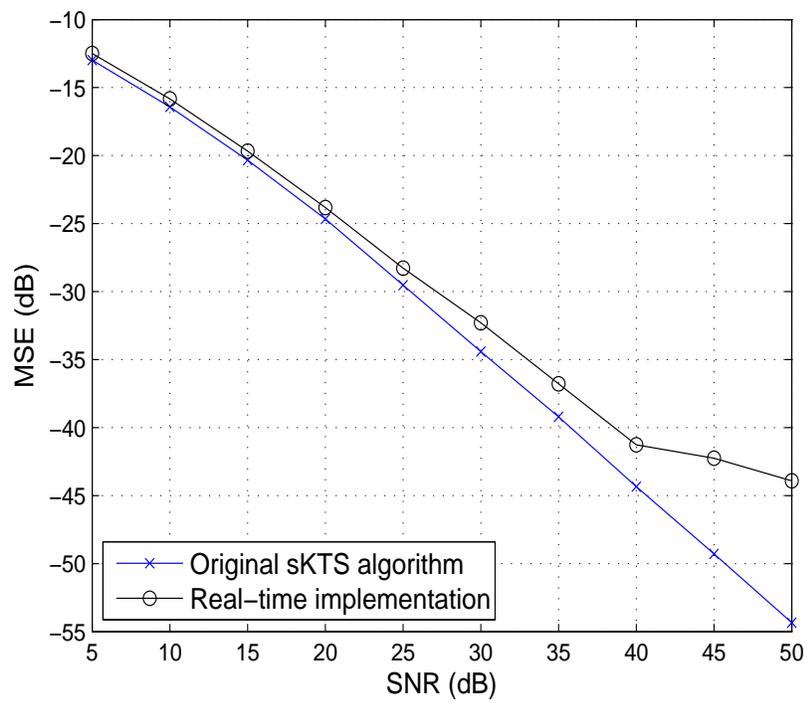, height=100mm, width=120mm}}
  \caption {The MSE performance of the RT-sKTS algorithm.}
  \label{fig:on}
\end{figure}

Next, we compare the performance of the RT-sKTS described in Section \ref{sec:sKTS_online} with the original sKTS algorithm. In this simulations, we set  $N=32$ and $T=30$. For the RT-sKTS algorithm, we set $\alpha = 0.4$. In order to test the performance in a harsh condition, we arbitrarily change the delay structure of the CIR for every 30 observation vectors. To ensure the convergence of  the online update strategy in (\ref{eq:rup1}) and (\ref{eq:rup2}), we use the first 10 observation vectors for  warming up purpose  and then use the rest for measuring the MSE performance.  Note that in practice, such warming up period would not be necessary since the support of channel vector would not be changed abruptly in many real applications.
In Fig.~\ref{fig:on}, we see that the RT-sKTS algorithm performs close to the original sKTS algorithm  in low and mid range SNR regime.
In the high SNR regime, however, the RT-sKTS algorithm suffers slight performance loss due to the approximation step of $\widetilde{\mathbf{d}}_{n}^{H}$ and $\widetilde{\mathbf{\Phi}}_{n}$. Nevertheless, as shown in Fig.~\ref{fig:on} and  Fig. \ref{fig:perf} (a),  the RT-sKTS algorithm maintains the performance gain over the conventional channel estimators.

%
\subsection{Experiments in Dynamic MRI Application}
\begin{figure} [!]
 \centering
\centerline{\epsfig{figure=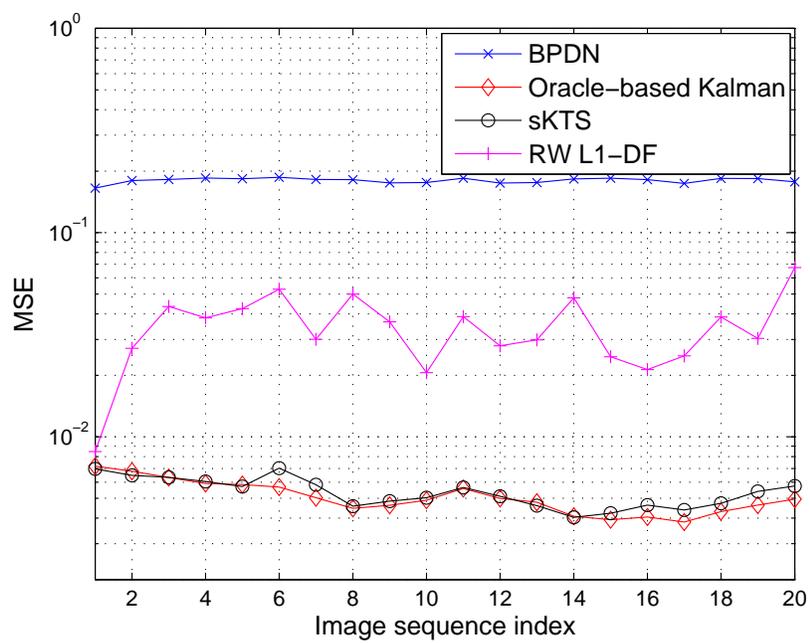,  width=120mm}}
  \caption {The MSE performance of the dynamic MRI image as a function of image sequence index.}
  \label{fig:cardiac}
\end{figure}

%
In this subsection, we investigate the performance of the sKTS algorithms in the reconstruction of the dynamic MRI images.
In our test, we use a sequence of $32 \times 32$ dimensional cardiac images  shown in Fig.~\ref{fig:cardiac}\footnote{These images are decimated from the original $128 \times 128$ images \cite{modifiedcs}. The raw image data is available online \cite{vaswani}.}. We generate the measurements by performing two dimensional discrete wavelet transform (DWT) with a 2-level Daubechies-4 wavelet, applying two dimensional DFT matrix and taking the $N$ randomly chosen frequency-domain image samples.
After adding the Gaussian noise to the image, we recover the original image using the recovery algorithms.
We set $N=358$, which corresponds to about 35\% of the image size (i.e., $M=1024$). We could empirically observe that the location of nonzero coefficients in wavelet image is slowly changing (i.e., support change occurs for only a few places), which matches well with our simultaneous sparse signal model.
In order to capture the most of signal energy, we set $K=152$ for all images\footnote{Actually, we set $K$ to the the number of coefficient containing $99.9$\% of the signal energy.}. In Fig.~\ref{fig:cardiac}, we plot the MSE of the several image recovery schemes obtained for each image. The sKTS algorithm outperforms the basis pursuit denoising (BPDN) \cite{candes} and RW1L-DF \cite{charles2} and also performs close to the Oracle-based Kalman smoother. Note that we could not include modified CS scheme in \cite{modifiedcs} in our numerical experiments since large number of measurement samples is required for the first image.

%
\section{Conclusions}
\label{sec:conclusion}
%
In this paper, we studied the problem to estimate the time-varying sparse signals when the sequence of the correlated observation vectors are available. In many signal processing and wireless communication applications, the support of sparse signals changes slowly in time and thus can be well modeled as simultaneously sparse signal, we proposed a new sparse signal recovery algorithm, referred to as sparse Kalman tree search (sKTS), that identifies the support of the sparse signal  using multiple measurement vectors. The proposed sKTS scheme performs the Kalman smoothing to extract the \emph{a posteriori} statistics of the source signals and the greedy tree search to identify the support of the signal. From the case study of sparse channel estimation problem in orthogonal frequency division multiplexing (OFDM) and image reconstruction in dynamic MRI, we demonstrated that the proposed sKTS algorithm is effective in recovering the dynamic sparse signal vectors.

\appendices
\section{Derivation of (\ref{eq:qqq})}
\label{appen:q}
From (\ref{eq:estep}) and (\ref{eq:vv2}), we get
\begin{align}
Q\left(\mathbf{c}_{i}; \hat{\mathbf{c}}_i^{(l)}\right) = & C' - \sum_{n=Ti+1}^{T(i+1)}E\left[  \frac{1}{\sigma_{w}^{2}} \left\|\mathbf{y}_{n} - \mathbf{B}_{n}{\rm diag}(\mathbf{c}_{i}) \mathbf{s}_{n} \right\|^{2} \bigg|  \mathbf{y}_{1:T} ; \hat{\mathbf{c}}_i^{(l)} \right]  \\
= & C'' + \frac{1}{\sigma_{w}^{2}} \sum_{n=Ti+1}^{T(i+1)}\bigg\{ E\left[{\rm tr} \left[2 {\rm Re}\left(  \mathbf{B}_{n} {\rm diag}(\mathbf{c}_{k}) \mathbf{s}_{n} \mathbf{y}_{n}^{H} \right)\right] \bigg| \mathbf{y}_{1:T} ; \hat{\mathbf{c}}_i^{(l)} \right]    \nonumber \\
& - E\left[{\rm tr}\left[ \mathbf{B}_{n} {\rm diag}(\mathbf{c}_{i}) \mathbf{s}_{n}\mathbf{s}_{n}^{H} {\rm diag}(\mathbf{c}_{i})  \mathbf{B}_{n}^{H}  \right]\bigg|  \mathbf{y}_{1:T} ; \hat{\mathbf{c}}_i^{(l)} \right] \bigg\}    \\
= & C'' + \frac{1}{\sigma_{w}^{2}} \sum_{n=Ti+1}^{T(i+1)}\bigg\{ {\rm tr} \left[2 {\rm Re}\left(  \mathbf{B}_{n} {\rm diag}(\mathbf{c}_{i}) E\left[\mathbf{s}_{n}\bigg| \mathbf{y}_{1:T} ; \hat{\mathbf{c}}_i^{(l)} \right] \mathbf{y}_{n}^{H} \right)\right]    \nonumber \\
& - {\rm tr}\left[ \mathbf{B}_{n} {\rm diag}(\mathbf{c}_{i})  E\left[ \mathbf{s}_{n}\mathbf{s}_{n}^{H}\bigg|  \mathbf{y}_{1:T} ; \hat{\mathbf{c}}_i^{(l)} \right] {\rm diag}(\mathbf{c}_{i})  \mathbf{B}_{n}^{H}  \right] \bigg\} ,   \\
\end{align}
where $C'$ and $C''$ are the terms independent of $\mathbf{c}_{i}$.
Using the property of the trace, i.e, ${\rm tr}(ABC) = {\rm tr}(BCA)  = {\rm tr}(CAB)$, we have
\begin{align}
Q\left(\mathbf{c}_{i}; \hat{\mathbf{c}}_i^{(l)}\right) =  & C'' + \frac{1}{\sigma_{w}^{2}} \sum_{n=Ti+1}^{T(i+1)} \bigg\{ 2 {\rm Re}\left(  \mathbf{y}_{n}^{H} \mathbf{B}_{n} {\rm diag}(\mathbf{c}_{i}) E\left[\mathbf{s}_{n}\bigg| \mathbf{y}_{1:T} ; \hat{\mathbf{c}}_i^{(l)} \right] \right) \nonumber \\
& - {\rm tr}\left[  \mathbf{B}_{n} {\rm diag}(\mathbf{c}_{i})  E\left[ \mathbf{s}_{n}\mathbf{s}_{n}^{H}\bigg|  \mathbf{y}_{1:T} ; \hat{\mathbf{c}}_i^{(l)} \right] {\rm diag}(\mathbf{c}_{i})  \mathbf{B}_{n}^{H}  \right]  \bigg\}
\end{align}

\section{Derivation of (\ref{eq:sterm})}
\label{appen:prf}
Denoting $\mathbf{b}_{n,i}$ as the transpose of the $i$th row vector of $\mathbf{B}_{n}$,  we can express the lefthand term of (\ref{eq:sterm}) as
\begin{align}
\mbox{left term} =   \sum_{j=1}^{M}    \mathbf{b}_{n,j}^{T} {\rm diag}(\mathbf{c}_{i}) \left(\Sigma_{n|1:T} + \widehat{\mathbf{s}}_{n|1:T}\widehat{\mathbf{s}}_{n|1:T}^{H} \right)  {\rm diag}(\mathbf{c}_{i}) {\rm conj}(\mathbf{b}_{n,j}).
\end{align}
Since $\mathbf{b}_{n,j}^{T} {\rm diag}(\mathbf{c}_{i}) = \mathbf{c}_{i}^{T} {\rm diag}(\mathbf{b}_{n,j})$ and ${\rm diag}(\mathbf{c}_{i}) {\rm conj}(\mathbf{b}_{n,j}) = {\rm diag}({\rm conj}(\mathbf{b}_{n,j}) ) \mathbf{c}_{i}$, we further have
\begin{align}
\mbox{left term} &= \sum_{j=1}^{M}   \mathbf{c}_{i}^{T}  {\rm diag}(\mathbf{b}_{n,j}^{T}) \left(\Sigma_{n|1:T} + \widehat{\mathbf{s}}_{n|1:T}\widehat{\mathbf{s}}_{n|1:T}^{H} \right)  {\rm diag}({\rm conj}(\mathbf{b}_{n,j})) \mathbf{c}_{i}  \\
 &=    \mathbf{c}_{i}^{T} \sum_{j=1}^{M} \left[ {\rm diag}(\mathbf{b}_{n,j}^{T}) \left(\Sigma_{n|1:T} + \widehat{\mathbf{s}}_{n|1:T}\widehat{\mathbf{s}}_{n|1:T}^{H} \right)  {\rm diag}({\rm conj}(\mathbf{b}_{n,j})) \right]\mathbf{c}_{i},
\end{align}
and hence we finally have
\begin{align}
\mbox{left term} =  \mathbf{c}_{i}^{T} \left({\rm conj}\left(\mathbf{B}_{n}^{H} \mathbf{B}_{n}\right)  \odot     \left(\Sigma_{n|1:T} + \widehat{\mathbf{s}}_{n|1:T}\widehat{\mathbf{s}}_{n|1:T}^{H} \right) \right) \mathbf{c}_{i}.
\end{align}

\clearpage

%



\clearpage

\clearpage

\begin{figure} [!]
 \centering
\centerline{\epsfig{figure=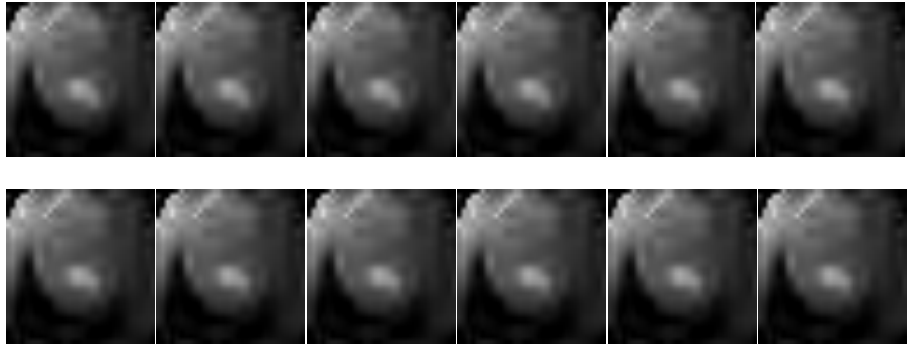,  width=120mm}}
  \caption {A sequence of cardiac MRI images.}
  \label{fig:image}
\end{figure}

\clearpage

\clearpage



\begin{thebibliography}{1}


\bibitem{candes} E. J. Candes, J. Romberg, and T. Tao, ``Robust uncertainty principles: exact signal reconstruction from highly incomplete frequency information,"  {\it IEEE Trans. Information Theory}, vol. 52, pp. 489-509, Feb. 2006.

\bibitem{omp} J. A. Tropp and A. C. Gilbert, ``Signal recovery from random measurements via orthogonal matching pursuit," {\it IEEE Trans. Information Theory}, vol. 53, pp. 4655-4666, Dec. 2007.

\bibitem{gomp} J. Wang, S. Kwon, and B. Shim, ``Generalized orthogonal matching pursuit," {\it IEEE Trans. Signal Process.}, vol. 60, pp. 6202-6216, Dec. 2012.

\bibitem{cosamp} D. Needell and J. A. Tropp, ``CoSaMP: iterative signal recovery from incomplete and inaccurate samples," {\it Appl. Comput. Harmon. Anal.}, vol. 26, pp. 301-321, 2009.

\bibitem{wei} W. Dai and O. Milenkovic, ``Subspace pursuit for compressive sensing signal reconstruction," {\it IEEE Trans. Information Theory}, vol. 55, pp. 2230-2249, May 2009.


\bibitem{rao_basis} D. P. Wipf and B. D. Rao, ``Sparse Bayesian learning for basis selection," {\it IEEE Trans. Signal Process.}, vol. 52, pp. 2153-2164, August 2004.

\bibitem{bcs} S. Ji, Y. Xue, and L. Carin, ``Bayesian compressed sensing," {\it IEEE Trans. Signal Process.}, vol. 56, pp. 2346-2356, June 2008.

%

\bibitem{tropp} J. A. Tropp, A. C. Gilbert, M. J. Strauss, ``Algorithms for simultaneous sparse approximation. Part I: greedy pursuit," {\it Signal Processing}, vol. 86, pp. 572-588, 2006.

\bibitem{rao_mmv} S. F. Cotter, B. D. Rao, K. Engan, K. Kreutz-Delgado, ``Sparse solutions to linear inverse problems with multiple measurement vectors," {\it IEEE Trans. Signal Process.}, vol. 53, pp. 2477-2488, July 2005.

\bibitem{tropp2}   J. A. Tropp, A. C. Gilbert, M. J. Strauss, ``Algorithms for simultaneous sparse approximation. Part II: convex relaxation," {\it Signal Processing}, vol. 86, pp. 589-602, 2006.


\bibitem{rao_emp} D. P. Wipf and B. D. Rao, ``An empirical Bayesian strategy for solving the simultaneous sparse approximation problem," {\it IEEE Trans. Signal Process.}, vol. 55, pp. 3704-3716, July 2007.

\bibitem{rao_temporal} Z. Zhang and B. D. Rao, ``Sparse signal recovery with temporally correlated source vectors using sparse Bayesian learning," {\it IEEE Journal of Selected Topics in Signal Processing}, vol. 5, pp. 912-926, Sept. 2011.

\bibitem{rao_ar} Z. Zhang and B. D. Rao, ``Sparse signal recovery in the presence of correlated multiple measurement vectors," {\it Proc. ICASSP 2010}, Dallas, TX USA, 2010, pp. 3986-3989.

\bibitem{prasad} R. Prasad, C. R. Murphy and B. D. Rao, ``Joint approximately sparse channel estimation and data detection in OFDM systems using sparse Bayesian learning," {\it IEEE Trans. Signal Process.}, vol. 62, no. 14, pp. 3591-3603, July 2014.


%


\bibitem{charles1}  A. S. Charles, M. Asif, J. Romberg, and C. Rozwell, ``Sparsity penalties in dynamical system estimation," {\it Proc. Conf. on Information Sciences (CISS)}, Mar. 2011, pp. 1-6.

\bibitem{charles2} A. S. Charles and C. J. Rozell, ``Re-weighted $\ell_1$ dynamic filtering for time-varying sparse signal estimation," {\it http:\\arxiv.org/abs/1208.0325}.

\bibitem{kalmancs} N. Vaswani, ``Kalman filtered compressed sensing," {\it Proc. ICIP } San Diego, CA USA, 2008, pp. 893-896.

\bibitem{schniter2} J. Ziniel and P. Schniter, ``Dynamic compressive sensing of time-varying signals via approximate message passing," {\it IEEE Trans. Signal Processing}, vol 61, pp. 5270-5284, Nov. 2013.

\bibitem{nicholas} N. Kalouptsidis, G. Mileounis, B. Babadi, and V. Tarokh, ``Adaptive algorithms for sparse system identification," {\it Signal Processing}, vol. 91, pp. 1910-1919, Aug. 2011.

\bibitem{giannakis} D. Angelosante, S. Roumeliotis, and G. Giannakis, ``Lasso-Kalman smoother for tracking sparse signals," {\it Asilomar conference}, Pacific Grove, CA USA, Nov. 2009, pp. 181-185.
		
\bibitem{jchoi} J. W. Choi, T. J. Riedl, K. Kim, A. C. Singer, and J. C. Preisig, ``Adaptive linear turbo equalization over doubly selective channels," {\it IEEE Journal of Oceanic Engineering}, vol. 36, pp. 473-489, Oct. 2011.			

\bibitem{em} A. P. Demster, N. M. Laird, and D. B. Rubin, ``Maximum likelihood from incomplete data via the EM algorithm," {\it J. R. Statist. Soc. B}, vol. 39, pp. 1-38, 1977.

\bibitem{doubly} J. W. Choi, K. Kim, T. J. Riedl, and A. C. Singer, ``Iterative estimation of sparse and doubly-selective multi-input multi-output (MIMO) channel," {\it Proc.  Signals, Systems and Computers Asilomar Conference}, Nov. 2009, pp. 620-624.




\bibitem{kailath} T. Kailath, A. H. Sayed, and B. Hassibi, {\it Linear estimation,} Prentice Hall, 2000.

\bibitem{bajwa} W. U. Bajwa, J. Haupt, A. M. Sayeed and R. Nowak, ``Compressed channel sensing: a new approach to estimating sparse multipath channels," {\it Proceedings of the IEEE}, vol. 98, pp. 1058-1076, June 2010.

\bibitem{berger} C. R. Berger, S. Zhou, J. C. Preisig and P. Willett, ``Sparse channel estimation for multicarrier underwater acoustic communication: from subspace methods to compressed sensing," {\it IEEE Trans. Signal Process.}, vol. 58, pp. 1708-1721, March 2010.

\bibitem{mp_cs} S. F. Cotter and B. D. Rao, ``Sparse channel estimation via matching pursuit with application to equalization," {\it IEEE Trans. Commun.}, vol. 50, pp. 374-377, March 2002.

\bibitem{akino} T. K. Akino, ``Optimum-weighted RLS channel estimation for rapid fading MIMO channels," {\it IEEE Trans. Wireless Comm.}, vol. 7, pp.4248-4260, Nov. 2008.

\bibitem {romano} M. B. Loiola, R. R. Lopes, and J. M. T. Romano, ``Modified Kalman filters for channel estimation in orthogonal space-time coded systems," {\it IEEE Trans. Signal Process.}, vol. 60, pp. 533-538, Jan. 2012



\bibitem{jake} W.\ C.\ Jakes, ``Microwave Mobile Communications,"
New York, John Wiley \& Sons Inc, 1975.


\bibitem{baddour} K. E. Baddour and N. C. Beaulieu, ``Autoregresive modeling for fading channel simulation," {\it IEEE Trans. Wireless Commun.}, vol. 4, pp. 1650-1662, July 2005.

\bibitem{lte101} 3GPP TS 36.101. ``User Equipment (UE) Radio Transmission and Reception." {\it 3rd Generation Partnership Project; Technical Specification Group Radio Access Network (E-UTRA)}. URL: http://www.3gpp.org.



\bibitem{vaswani} http://www.ece.iastate.edu/$\sim$namrata/software.html.

\bibitem{modifiedcs} N. Vaswani and W. Lu, ``Modified-CS: modifying compressive sensing for problems with partially known support," {\it IEEE Trans. Signal Process.}, vol. 58, no. 9, pp. 4595-4607, Sept. 2010.




%
%
%
%



\end{thebibliography}
\end{document}